\documentclass{asl} 

\pdfoutput=1

\usepackage{amsfonts}
\usepackage{amssymb}
\usepackage{latexsym}
\usepackage{eurosym}
\usepackage{epsfig}   
\usepackage{color}
\usepackage{ifthen}
\usepackage{verbatim}
\usepackage[numbers, sort&compress]{natbib}
\usepackage[all]{xy}
 
\def\bR{\begin{color}{red}} 
\def\bB{\begin{color}{blue}}
\def\bM{\begin{color}{magenta}}
\def\bC{\begin{color}{cyan}}
\def\bW{\begin{color}{white}}
\def\bBl{\begin{color}{black}} 
\def\bG{\begin{color}{green}}
\def\bY{\begin{color}{yellow}}
\def\e{\end{color}}

\theoremstyle{definition}
\theoremstyle{definition}

\newcommand{\bit}{\begin{itemize}}
\newcommand{\eit}{\end{itemize}\par\noindent}
\newcommand{\ben}{\begin{enumerate}}
\newcommand{\een}{\end{enumerate}\par\noindent}
\newcommand{\beq}{\begin{equation}}
\newcommand{\eeq}{\end{equation}\par\noindent}
\newcommand{\beqa}{\begin{eqnarray*}}
\newcommand{\eeqa}{\end{eqnarray*}\par\noindent}
\newcommand{\beqn}{\begin{eqnarray}}
\newcommand{\eeqn}{\end{eqnarray}\par\noindent}

\newcommand{\jump}{\par\vspace{2.0mm}\par\noindent}

\theoremstyle{plain}
\newtheorem{thm}{Theorem}

\title[The logic of quantum mechanics -- Take II]{The logic of quantum mechanics -- Take II}
\author{Bob Coecke}
\revauthor{Coecke, Bob}
\address{University of Oxford,\\ Department of Computer Science,\\ Quantum Group} 
\email{coecke@cs.ox.ac.uk}
\thanks{The work presented here is supported by the British Engineering and Physical Research Council (EPSRC), the US Office of Naval Research (ONR) and the Foundational Questions Institute (FQXi).  The content of this paper reflects a series of seminars in 2010--2012 with as titles: ``Monoidal categories as an axiomatic foundation'', ``In the beginning God created tensor ... then matter ... then speech'', ``How computer science helps bringing quantum mechanics to the masses'', ``Selling categories to the masses'', or the actual title of this paper itself.}

\begin{document}
\maketitle

\begin{abstract}
We put forward a new take on the logic of quantum mechanics, following Schr\"odinger's point of view that it is composition which makes quantum theory what it is, rather than its particular propositional structure due to the existence of superpositions, as  proposed by Birkhoff and von Neumann.  This gives rise to an intrinsically quantitative kind of logic, which truly deserves the name `logic' in that it also models meaning in natural language, the latter being the origin of logic,  that it supports automation, the most prominent practical use of logic, and that it  supports probabilistic inference.
\end{abstract}

\section{The physics and the logic of quantum-ish logic}


In 1932 John von Neumann formalized Quantum Mechanics in his book  ``Mathematische Grundlagen der Quantenmechanik''.  This was  effectively the official birth of the quantum mechanical formalism which until now, some 75 years later, has remained the same.  Quantum theory  underpins so many things in our daily lives including  chemical industry, energy production and information technology, which arguably makes it the most technologically successful theory of physics ever.  

However,  in 1935, merely three years after the birth of his brainchild,  von Neumann wrote in a letter to American mathematician Garrett Birkhoff :``I would like to make a confession which may seem immoral: I do not believe absolutely in Hilbert space no more.'' (sic)---for more details see \cite{Redei}.  

Soon thereafter they published a paper entitled ``The  {\em Logic}  of Quantum Mechanics''  \cite{BvN}.  Their `quantum logic' was casted in order-theoretic terms, very much in the  spirit of the then reigning algebraic view of logic, with the distributive law being replaced with a weaker (ortho)modular law.

This resulted in a research community of \em quantum logicians \em \cite{Mackey,Piron,Foulis,CMW}.  However, despite von Neumann's reputation, and the large body of research that has been produced in the area, one does not find a trace of this activity neither in the mainstream physics,  mathematics, nor   logic literature. Hence, 75 years later one may want to conclude that this activity was a failure.  

What went wrong?

\subsection{The mathematics of it}

Let us consider the raison d'\^etre for the Hilbert space formalism.  So why would one need all this `Hilber space stuff', i.e.~the continuum structure, the field structure of complex numbers, a vector space over it, inner-product structure, etc. Why?  According to von Neumann, he simply used it because it happened to be `available'.  The use of linear algebra and complex numbers in so many different scientific areas, as well as results in model theory, clearly show that quite a bit of modeling can be done using Hilbert spaces.   On the other hand, we can also model any movie by means of the data stream that  runs through your cables when watching it.  But does this mean that these data streams make up the stuff that makes a movie?  Clearly not, we should rather turn our attention to the stuff that is being taught at drama schools and directing schools.  Similarly, von Neumann turned his attention to the actual physical concepts behind quantum theory, more specifically, the notion of a physical property and the structure imposed on these by the peculiar nature of quantum observation.  His quantum logic gave the resulting `algebra of physical properties' a privileged role.  All of this leads us to ... 

\subsection{... the physics of it}

Birkhoff and von Neumann crafted quantum logic in order to emphasize the notion of quantum \em superposition\em.  In terms of states of a physical system and properties of that system,  superposition means that the strongest property which is true for two distinct states  is also  true for states other than the two given ones. In order-theoretic terms this means, representing states by the atoms of a lattice\footnote{I.e.~a partially ordered set with a minimal element $0$ and maximal element $1$, and in which each pair of elements has a supremum and an infimum.  In fact, there are physical resons for assuming that this lattice is \em complete \em \cite{Piron,DJMoore}, i.e.~arbitrary suprema and infima exist.} of properties \cite{DJMoore}, that the join $p\vee q$ of two atoms $p$ and $q$ is also above other atoms.  From this it easily follows that the distributive law\footnote{\em Distributivity \em means that for any elements $a, b, c$ of the lattice we have that $a \wedge ( b\vee c)=(a\wedge b)\vee(a\wedge  c)$ and that $a \vee ( b\wedge c)=(a\vee b)\wedge(a\vee  c)$.} breaks down: given atom\footnote{An \em atom \em is an element $p\not= 0$ which is such that whenever $a<p$ then $a=0$.} $r\not= p, q$ with $r< p\vee q$ we have $r \wedge ( p\vee q)=r$ while $(r\wedge p)\vee(r\wedge  q)= 0\vee 0 = 0$.  Birkhoff and von Neumann as well as  many others believed that understanding the deep structure of superposition is the key to obtaining a better understanding of quantum theory as a whole.  But as already mentioned, 75 years later quantum logic did not break through.  

The Achilles' heel of quantum logic is the fact that it fails  to elegantly capture  `composition of quantum systems', that is, how do we describe multiple quantum systems given that we know how to describe the individual quantum systems.  On the other hand, also in 1935, Schr\"odinger pushed forward the idea that the stuff which truly characterizes quantum behavior is precisely the manner in which  quantum systems compose \cite{Schrodinger}.  Over the past 30 years or so we have seen ample evidence for this claim.  So-called `quantum non-locality' was experimentally confirmed, and the focus on quantum information processing has revealed a wide range of quantum phenomena which all crucially depend  on the manner in which quantum systems compose, most notably exponential quantum computational speed-up which led to the quantum computing paradigm \cite{Shor}.

Now reversing the roles, rather than explaining all of quantum theory in terms of superposition, can  we maybe  explain all of quantum theory in term of the manner in which quantum systems compose, including superposition? 

\subsection{The game plan}  

Here is the list of tasks we've set ourselves:
\bit
\item {\bf Task 0.} First we want to solve:
\[
{\mbox{\rm tensor product structure} \over \mbox{\rm the other Hilbert space stuff}} = {\bf\Huge???}
\]
that is, we want to know what remains of the Hilbert space formalism if we `remove all of its structure except for the manner in which  systems compose'.  In other words, we want to axiomatize composition of systems, which we denote by $\otimes$, without  any reference to underlying spaces.
\item {\bf Task 1.}  Next we investigate which additional assumptions on $\otimes$ are needed in order to deduce experimentally observed  phenomena?  That is, given that the structure deduced in Task 0 applies to a wide range of theories (as we shall see below in Section \ref{sec:proclogic})  what extra structure do we need to add such that the resulting framework allows us to derive typical quantum behaviors.
\item {\bf Task 2.}  Once this `typically quantum' structure has been identified, we take on the challenge to find this same structure elsewhere in what we usually conceive as  `our classical reality'.  This may involve looking at this classical reality through a `novel pair of glasses'.
\eit
And, ... here are the resulting outcomes:
\bit
\item {\bf Outcome 0:} 
That was an easy one.  The solution to this has been around for quite a while.  It is called \em symmetric monoidal category \em \cite{Benabou}.  In fact, as discused in \cite{ContPhys,CatsII}, physical processes themselves form a \em strict \em symmetric monoidal category, while set theory based models such as the Hilbert space model are typically non-strict, which invokes so-called `coherence conditions' \cite{MacLaneCoherence} between `natural transformations' \cite{EilenbergMacLane}.  But one can show that an arbitrary symmetric monoidal category is always `categorically equivalent' to a strict symmetric monoidal category, which means that, up to  isomorphisms, whatever one can do with a non-strict one, one can do with a strict one too. Hence, here we will only spell out strict symmetric monoidal categories, in terms of their \em graphical language \em \cite{Penrose,JS}, that is, a language which is such that an equational statement  holds in it if and only if it follows from the axioms of a strict symmetric monoidal category.
\item {\bf Outcome 1a:}  Quoting Princeton philosopher Hans Halvorson in his editorial to the volume \em Deep Beauty: Understanding the Quantum World through Mathematical Innovation \em which marked 75 years since the publications of von Neumann's quantum formalism \cite{HalvorsonBook}: ``What is perhaps most striking about Coecke's approach is the sheer ratio of results to assumptions.''  As we shall see below, with very little additional structure one can already derive a wide range of quantum phenomena, and the required computations are utterly trivial.  This is in sharp contrast with Birkhoff-von Neumann quantum logic where one couldn't derive much; and in the case that one could derive something physically relevant one had to work really hard.
\item {\bf Outcome 1b:} Moreover, exposing this structure  has already helped to solve standing open problems in quantum information, e.g.~\cite{DP2,Clare,BoixoHeunen}, and provided novel insights in the nature of quantum non-locality \cite{CES,CDKZ}.  
\item {\bf Outcome 1c:} The diagrammatic framework underpinning strict symmetric monoidal categories has meanwhile been adopted by several leading researchers in quantum foundations e.g.~\cite{Chiri1,Chiri2,HardyPicturalism}; quoting Lucien Hardy in \cite{HardyPicturalism}:
``[...] we join the {\em quantum picturalism\em} revolution \cite{ContPhys}''.
\item {\bf Outcome 2a:} Observe the following similar looking pictures:
\begin{center}
\epsfig{figure=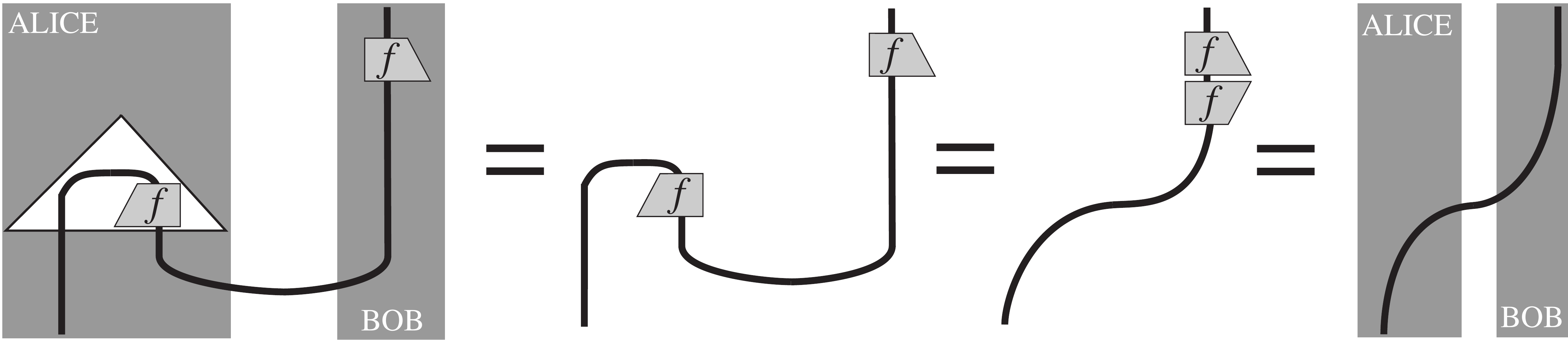,width=320pt}
\end{center}
\jump
\begin{center}
\epsfig{figure=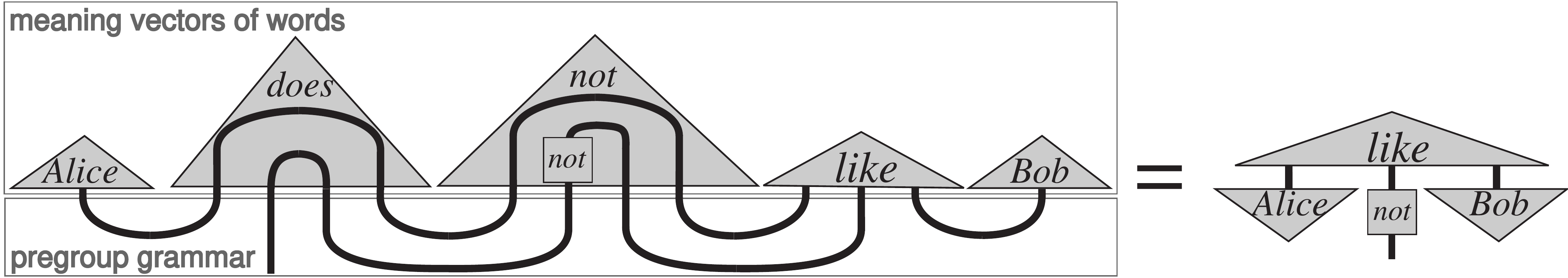,width=320pt}
\end{center}
\begin{center}
\epsfig{figure=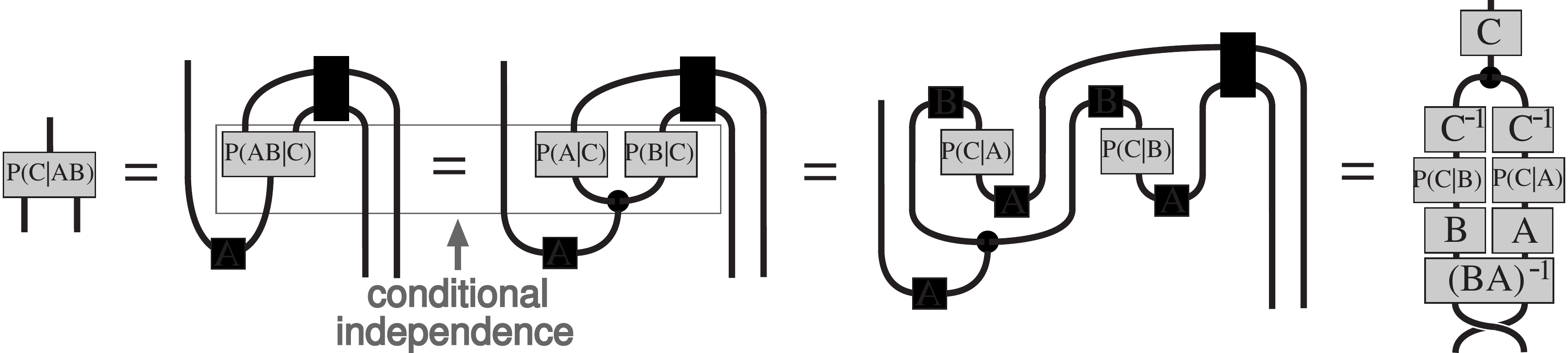,width=320pt}
\end{center}
These are respectively taken from a physics paper on the flow of information in quantum protocols \cite{AC1,Kindergarten,ContPhys}, a linguistics paper on how to compute the meaning of a sentence given the meaning of its words \cite{CCS,CSC}, and a probability theory paper that axiomatizes Bayesian inference \cite{CSBayes}.  The graphical calculi are in each case very similar, which points at a common reasoning systems in each of these very distinct areas.  Note in particular that in each case the  data of interest is of a fundamentally quantitative  nature.  Could this be pointing at the existence of some sort of \em quantitative logic\em, which is not typical to these areas but of a more universal nature?
\eit
So let us now consider  ...  

\subsection{... the logic of it}

\begin{figure}
\centering
\epsfig{figure=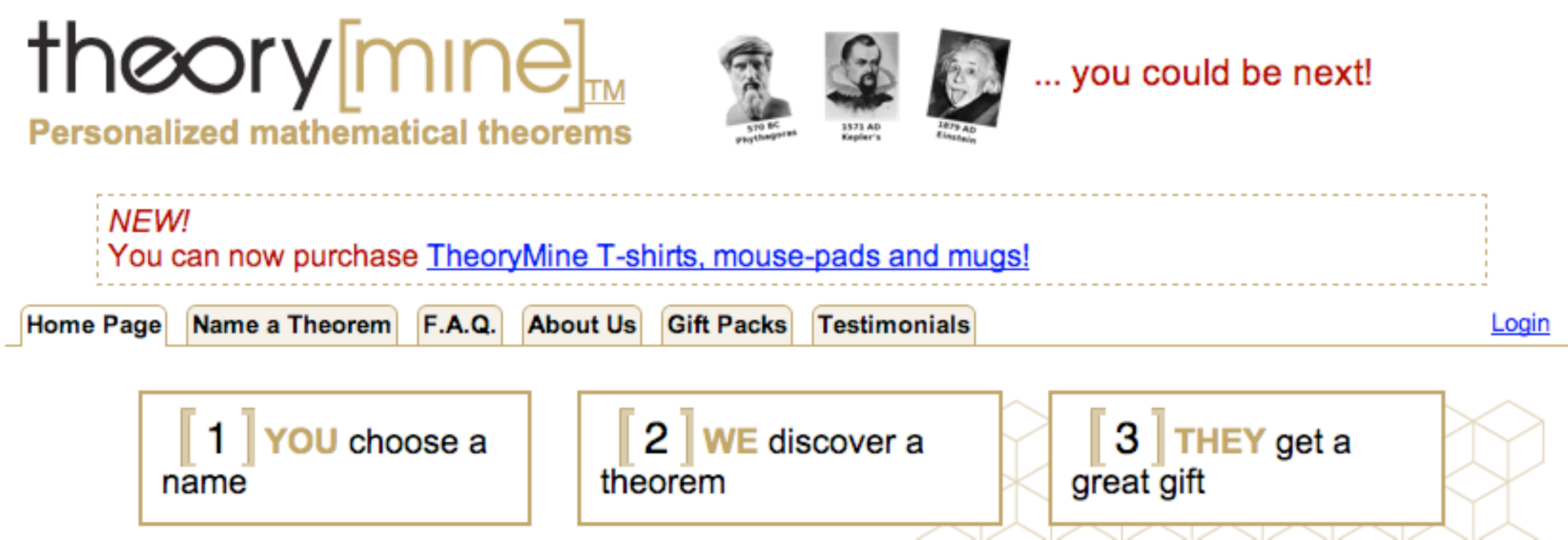,width=280pt}
\caption{The theory[mine] website which allows one to buy an automatically generated theorem and name it after someone.  It is a novelty gift spin-off from the automated theory exploration expertise at Edinburgh University---see \cite{TheoryMine} for the science.}\label{fig:TheoryMine}
\end{figure}

\begin{figure}
\centering
\epsfig{figure=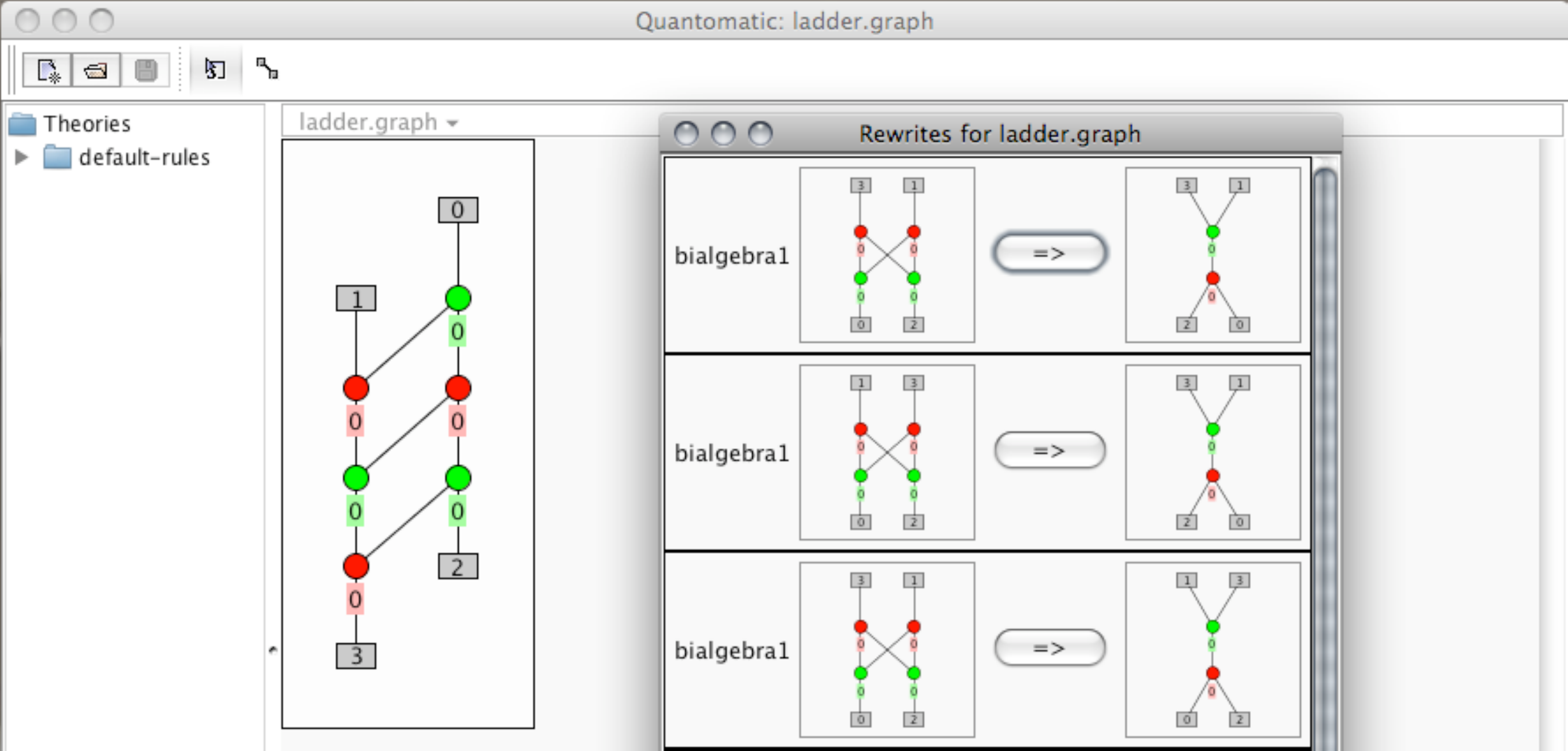,width=280pt}
\caption{Screenshot of the quantomatic software  developed in a collaboration between Oxford and Google, which can be downloaded from http://sites.google.com/site/quantomatic/.}  \label{fig:Quantosite}
\end{figure}

What is logic? The previous century has known a huge proliferation of logics of various kinds, and there probably are as many opinions of what logic actually is. Rather than making a case for one or another logical paradigm we will take a pragmatic stance and conceive logic in terms of its origin and its most prominent practical use:
\bit
\item {\bf Origin: structure in natural language.} The origin of logic, tracing back to Aristotle, is that it is about `arguments in natural language'.   Consider for example the sentence:``Alice  and Bob either ate  everything  or nothing, then got sick.''  By using connectives, quantifiers, variable $f$ referring to food, constants $a$(lice) and $b$(ob), and predicates $Sick({\tt person})$ and $Eat({\tt person}, {\tt some\ kind\ of\ food})$    we can formalize this as follows:
\[
\,\ \ \ \ \ \ (\forall f:  Eat(a,\! f) \wedge Eat(b,\!  f))\vee \neg (\exists f:  Eat(a,\!  f) \wedge Eat(b,\!  f))  \Rightarrow Sick(a), Sick(b)
\]
However, statements like this are still tightly related to a truth-concept, that is, we classify statements in terms of these either being true or not.  Clearly there is a lot more to the \em meaning \em of a sentence than it either being true or false.  This leads us to the following questions: What do we mean by meaning? What is the logic governing meaning, more specifically,  how do meanings of words interact to form meanings of sentences?
\item {\bf Use: automated reasoning.}  Logic now forms the foundation for fields like automated proof checking and automated theorem proving in computer science, which are key to modern methods for verifying the correctness of new software and hardware.  Logic also controls robot behaviors in  artificial intelligence.   Even more adventurous is \em automated theory exploration\em, where one does not only try to automatically proof theorems, but also generate them, which is a much harder task (cf.~P vs.~NP)---see also Figure \ref{fig:TheoryMine}. 
\eit
Our diagrammatic framework appeals to both of these senses of logic, and in doing so produced important  new applications in each of these areas: 
\bit
\item The above depicted framework for modeling how  meanings of words interact to form meanings of sentences, introduced by  Clark, Sadrzadeh and the author  in \cite{CCS,CSC}, is the first to do so based on a clear conceptual underpinning.  It was a cover heading feature in New Scientist \cite{NewScientist} and meanwhile 
greatly improved performance of several natural language processing tasks   \cite{GrefSadr}.  We explain this framework in Section \ref{sec:compdistmeaning}, as well as its structural relationship the graphical quantum formalism. 
\item The diagrammatic formalism underpins the automated reasoning software  {\tt quantomatic} developed at Oxford and Google by   Dixon, Kissinger, Merry, Duncan, Soloviev and Frot---see also Figure \ref{fig:Quantosite}.  More recently, work on automated theory exploration of graphical theories  also started at Oxford \cite{QuantoCosy}, building further on the work done at Edinburgh \cite{IsaCosy}.  We won't discuss this here; details are in  \cite{DD1,DK,quanto} and on the {\tt quantomatic} website.
\eit

\section{Minimal process logic}\label{sec:proclogic} 

By a process logic we simply mean any strict symmetric monoidal category, and by minimal we mean that at this stage we consider no structure (yet) other than the strict symmetric monoidal structure.  We explain this structure in terms of its graphical language.  

We could as well have given a symbolic presentation. We refer the reader to \cite{ContPhys,CatsII} for such a symbolic presentation, exemplified for the specific case of cooking processes, and how they compose to make up recipes---\cite{ContPhys} also discusses how  a process logic explains why tigers have stripes while lions don't.

\subsection{Graphical language}

The data of a minimal process logic consists of processes, represented by boxes, each of which takes some type of systems as its input, represented by (an) input wire(s), and some type of system as its output:
\begin{center}
\quad\qquad\epsfig{figure=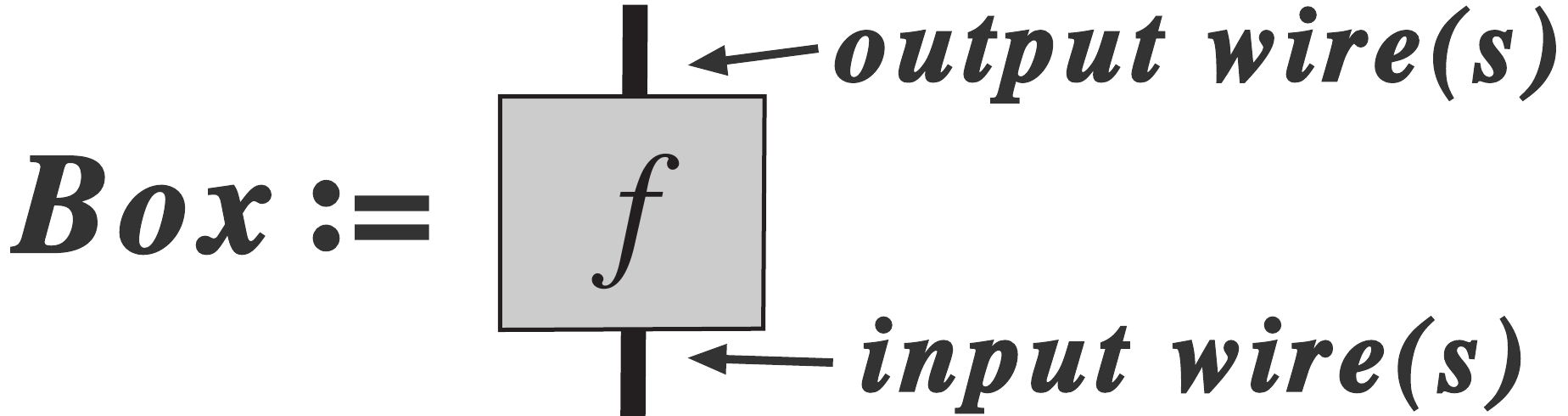,width=120pt}
\end{center}
These types may be compound, or trivial, i.e.~representing `no system':
\begin{center}
\ one system\    \quad $n$ \em sub\,\em-systems  \  \ \ \ \  \em no \em system \ \
\end{center}
\[
\underbrace{\epsfig{figure=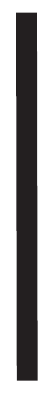,width=4pt}}_1\quad\ \ \ \qquad\underbrace{\epsfig{figure=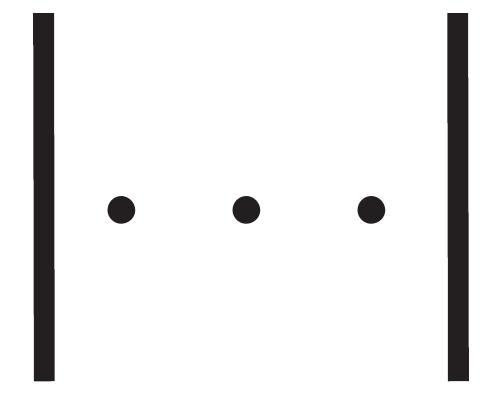,width=40pt}}_n\qquad\ \ \quad\underbrace{\epsfig{figure=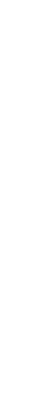,width=4pt}}_0
\]
Examples of types could be a particular quantum system, classical data of a certain size, grammatical types, e.g.~the type of a noun, verb, or a sentence, etc.  A process with no input wire is called a \em state\em---one can think of these as `preparation processes'.  Those with neither an input type nor an output type are called  \em values\em.  A process without an output type is called a \em valuation\em. 

The connectives of a minimal process logic constitute composition of processes. There are two modes: \em sequential \em or \em causal \em or \em connected \em composition, and, \em parallel \em or \em acausal \em or  \em disconnected \em composition, respectively depicted as:
\[
\raisebox{-1.02cm}{\epsfig{figure=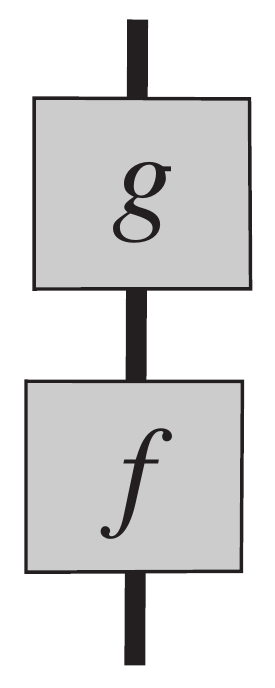,width=24pt}}
\qquad\qquad\qquad\qquad
\raisebox{-0.52cm}{\epsfig{figure=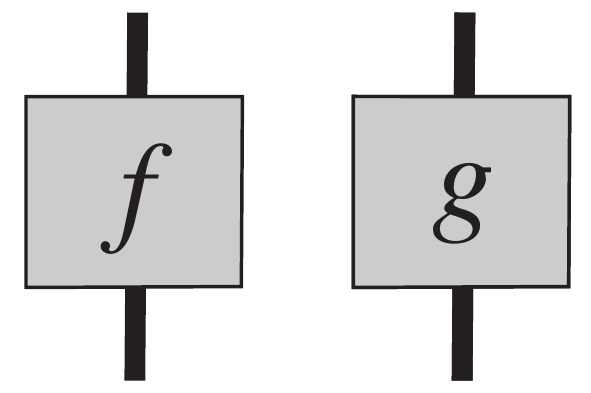,width=52pt}}
\]
So by post-composing a state with a valuation one obtains a value.  Note that sequential composition requires the output type of $f$ to be equal to the input type of $g$ while no such restriction exists for parallel composition.

\begin{figure}
\centering
\centerline{\epsfig{figure=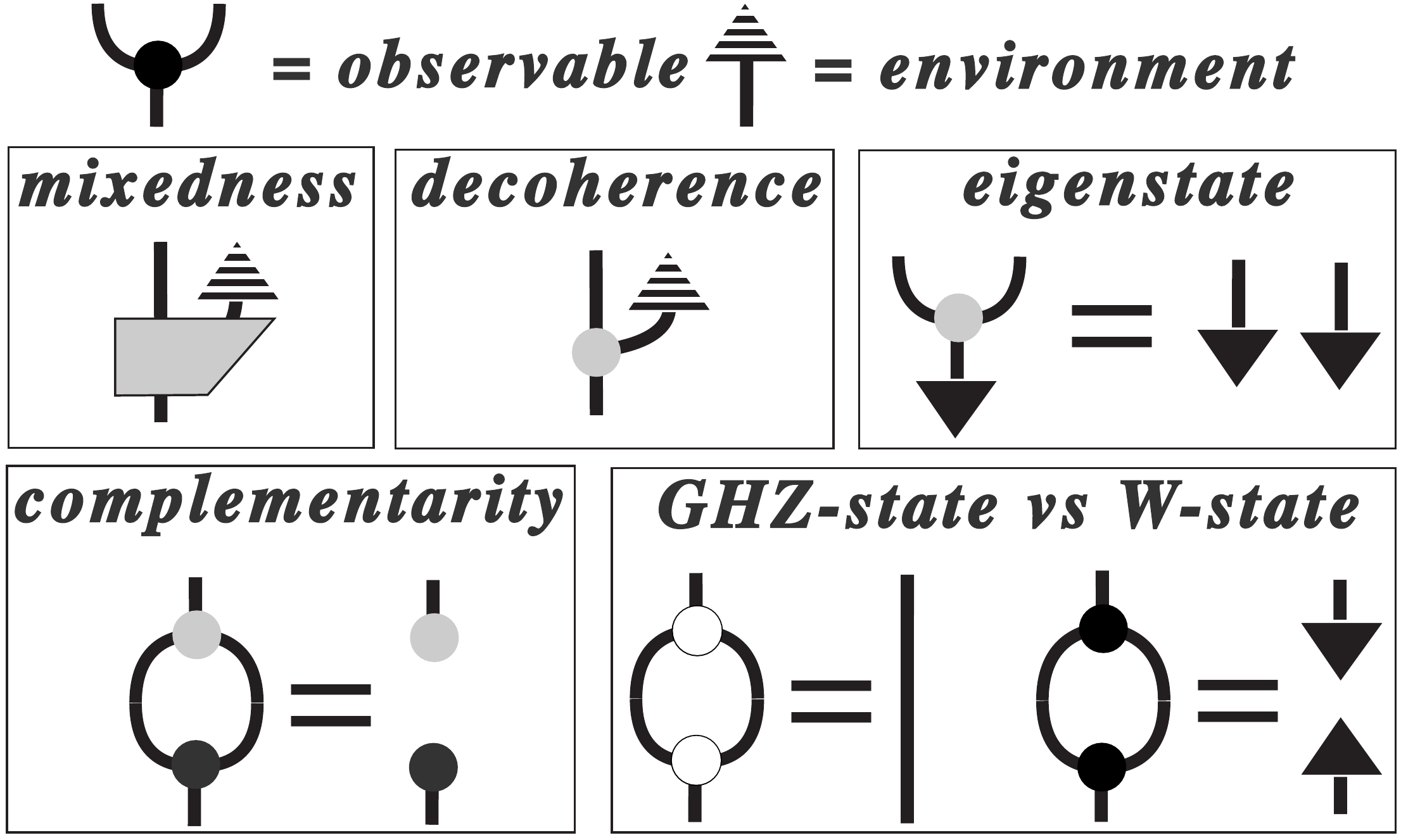,width=170pt}}
\caption{Examples of quantum mechanical concepts that can be expressed in purely topological terms, with the help of some new graphical elements.  They are taken from \cite{SelingerAxiom,CPer,CPav,CD2,CK}.}  \label{fig:topology}
\end{figure}

The formal paradigm underpinning minimal process logic is a topological one:
\begin{center}
\epsfig{figure=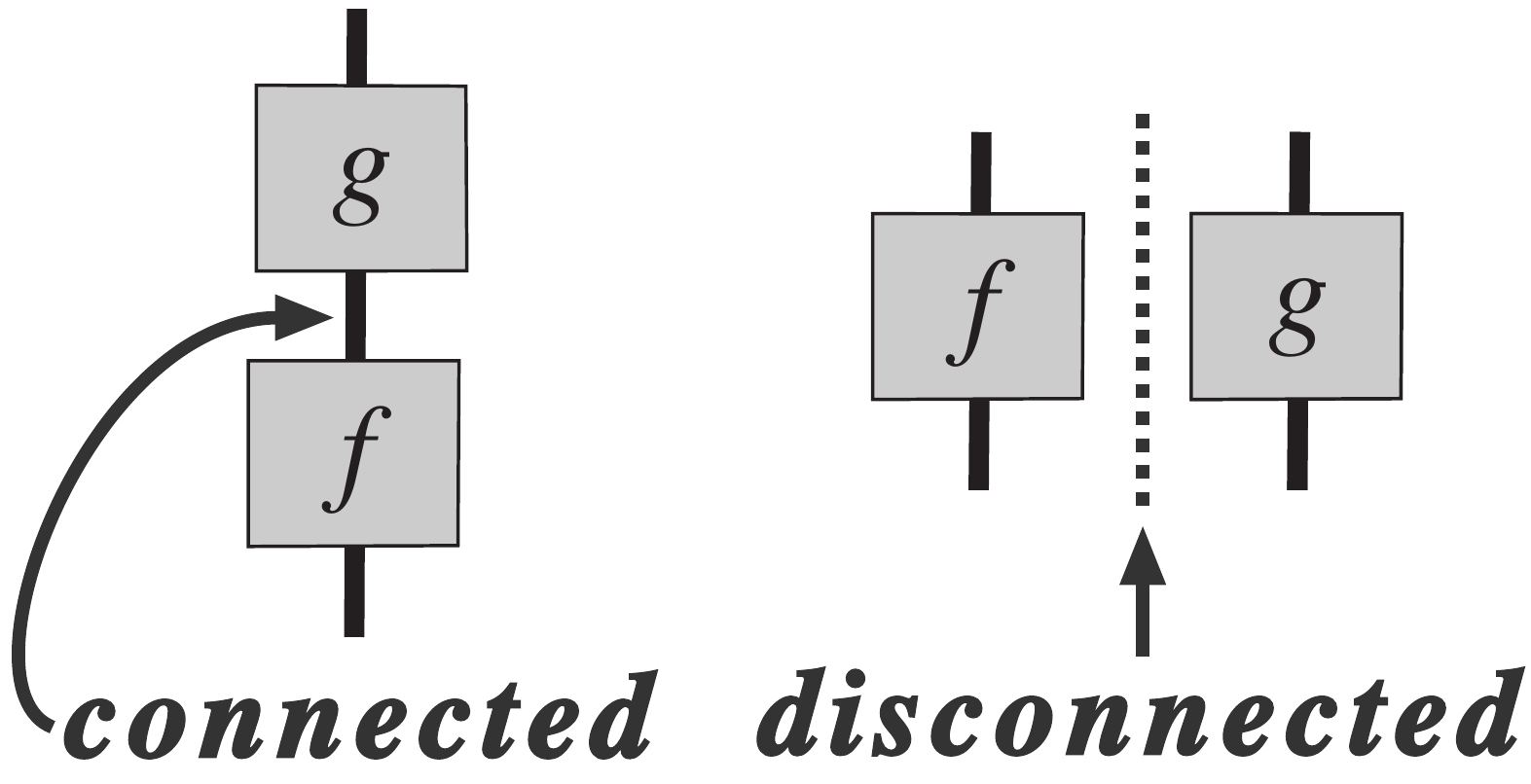,width=135pt}
\end{center}
The topology captures `what interacts with what', a wire standing for interaction while no wire stands for no interaction.  It is surprising how much concepts can be expressed purely in these topological terms---e.g.~see Figure \ref{fig:topology} for some topologically characterized quantum mechanical concepts.

The computational content of minimal process logic boils down to the simple intuitive rule that topologically equivalent diagrams are equal.  Hence computation proceeds by topological deformations:
\begin{center}
\epsfig{figure=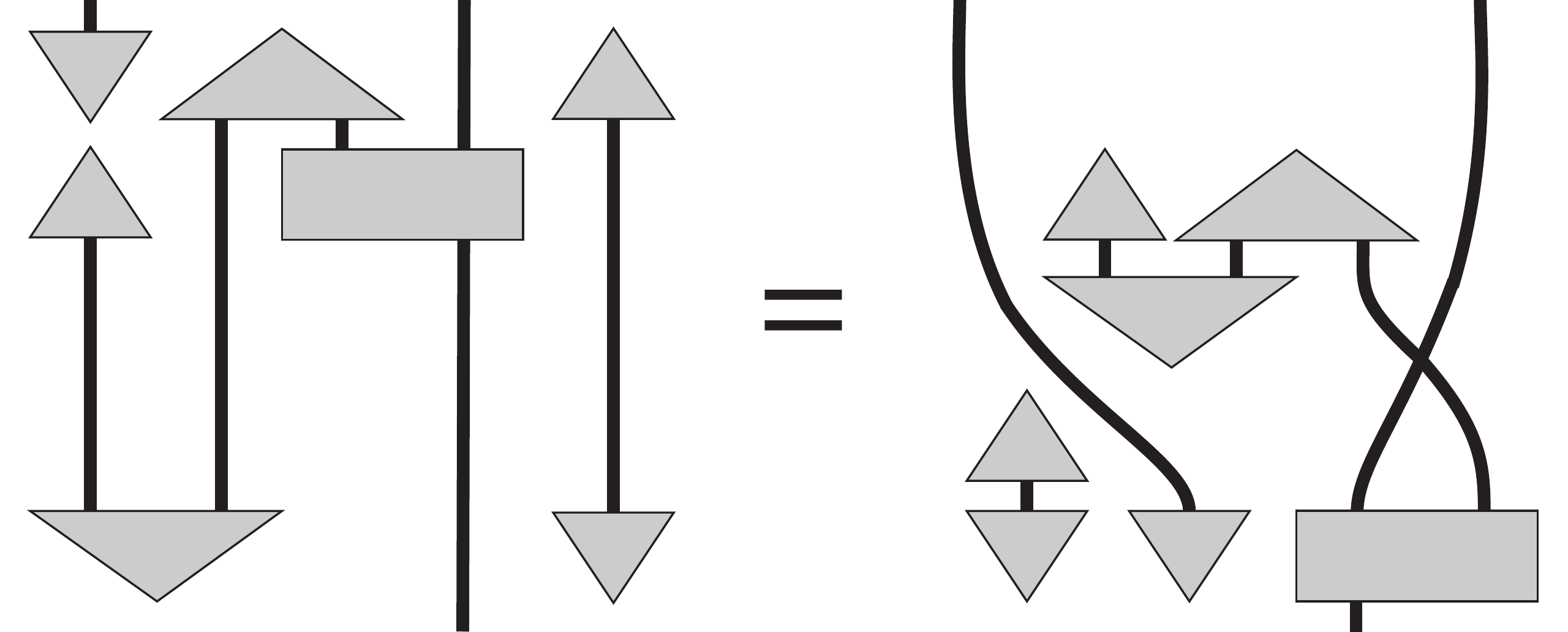,width=200pt}
\end{center}

There is no additional equational content to a minimal process logic.  This may sound surprising, since a strict symmetric monoidal category is subject to a number of axioms.  The explanation is that in the graphical language all these equations become tautologies.  For example, denoting sequential composition by $\circ$ and  parallel composition by $\otimes$, the `bifunctoriality equation' $(g\circ f)\otimes(k\circ h)=(g\otimes k)\circ (f\otimes h)$ of monoidal categories  becomes:
\begin{center}
\epsfig{figure=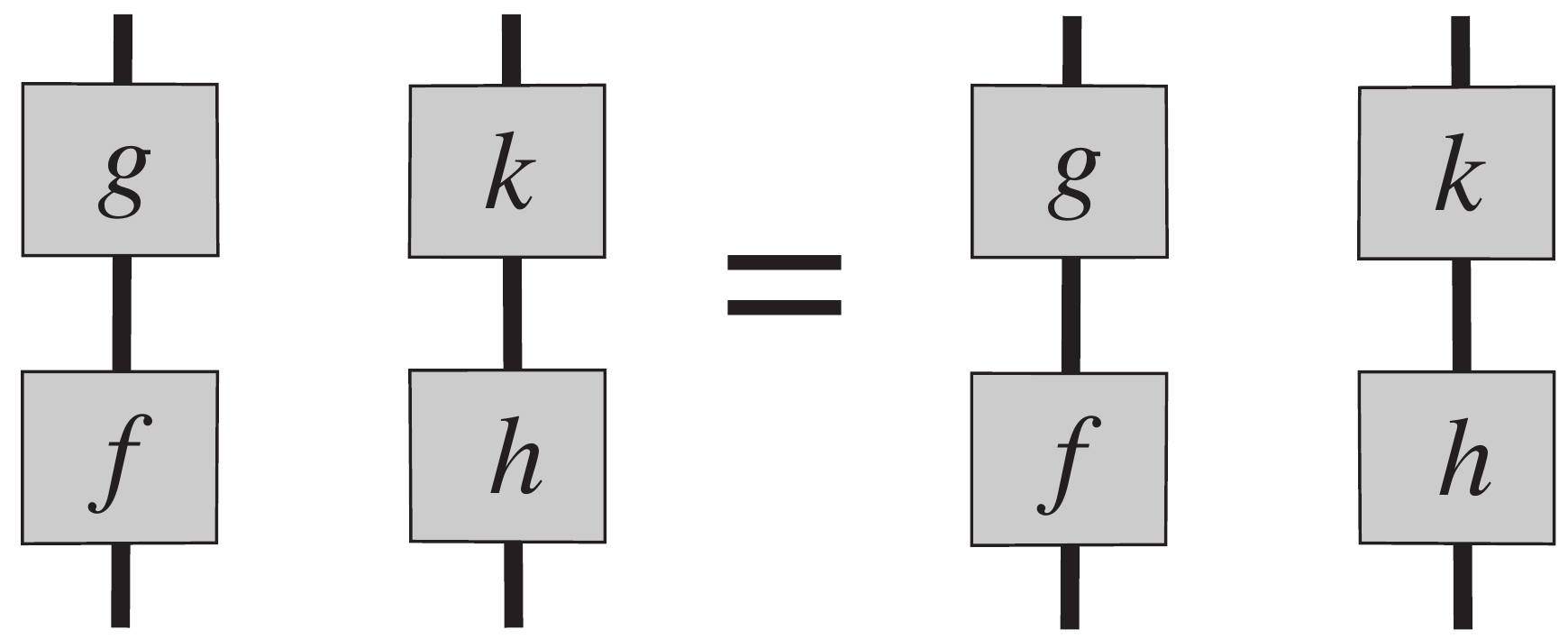,width=150pt}
\end{center}
In terms of processes this means that `$g$ after $f$, while, $k$ after $h$' is the same as `$g$ while $k$, after, $f$ while $h$'.  

\section{Quantum process logic - Take IIa}

Our next goal is to derive some non-trivial quantum phenomena by endowing a minimal process logic with a tiny bit of  extra structure, identified by Abramsky and the author in \cite{AC1, AC2}. 

\subsection{Dagger compact structure}

The first bit of extra structure will induce some kind of  metric on the states, namely, we will ask that each state can be turned into a valuation; applying this valuation to any other state will yield a value. Note that this is exactly how the highly successful Dirac notation \cite{Dirac} works: a ket $|\psi\rangle$  can be turned into a bra $\langle \psi|$, and when composing $\langle \psi|$ with another ket $|\phi\rangle$ we obtain a bra-ket $\langle \psi|\phi\rangle$ i.e.~an inner-product.  Since states may themselve arise by composing processes other than states, we will allow for the inputs and the outputs of any process to be `flipped':
\[
\forall\ \raisebox{-0.64cm}{\epsfig{figure=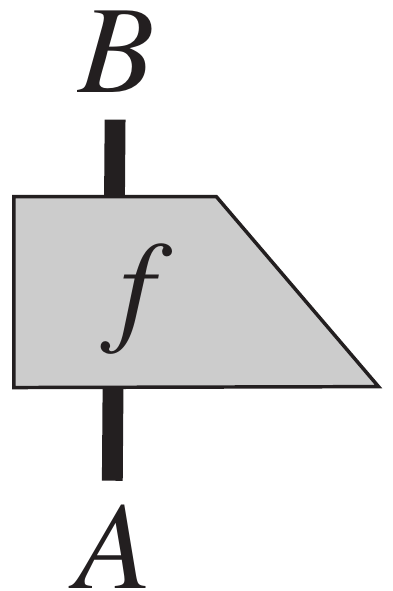,width=30pt}}\ \exists !\ \raisebox{-0.64cm}{\epsfig{figure=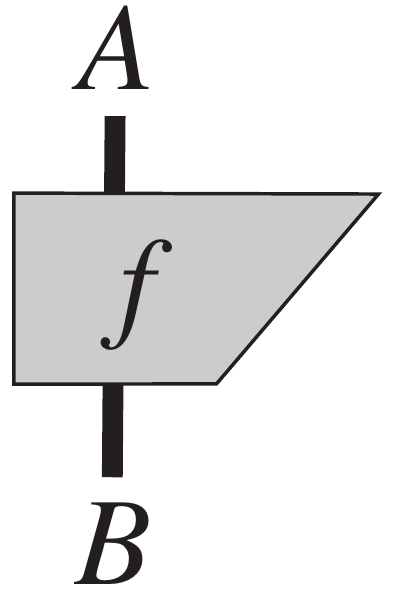,width=30pt}} 
\]
Note that flipping twice yields the original box, so flipping is involutive, and it is also clear that it preserves parallel composition, while it reverses sequential composition.  We refer to flipping as the \em adjoint \em or \em dagger\em.\footnote{From the perspective of Birkhoff-von Neumann quantum logic, one could conceive this as the analog to an orthocomplementation  on the lattice structure. That is, an order-reversing involution. Note in particularly that for non-Boolean lattices an orthocomplementation is a structure, not a property, as there can exist many different ones on the same lattice. In lattice theoretic terms the linear algebraic adjoint indeed arises as an expression involving Galois adjoints $(-)^*$and orthocomplementation $(-)'$, namely $f^\dagger(a)= (f^*(a'))'$ \cite{FMC,CMoore}.}

So far we haven't said anything specific about the parallel composition.  Now we will truly follow Schr\"odinger's path and specify in which manner quantum systems compose differently than classical systems.  In other words, we will assert that pure quantum states admit entanglement, diagrammatically:
\[ 
{{{quantum}\over{classical}} 
\ =\  
{\epsfig{figure=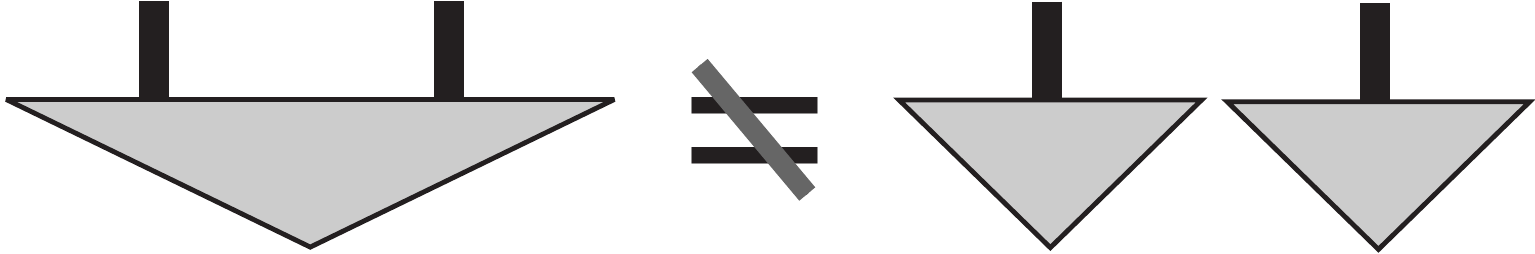,width=105pt}\over\epsfig{figure=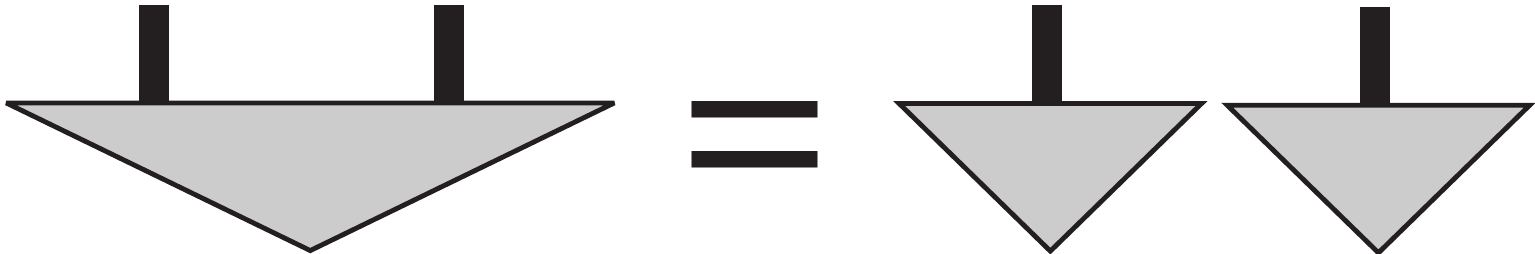,width=105pt}}}
\]
That is, a quantum state of two systems can in general not be described by describing the state of its parts.  Note that this is also not the case for probabilistic classical  data: a situation of two systems which comes with the promise that the states of the system are the same but unknown, can also not be described by independently describing the state of each system.  However, in quantum theory this  already occurs for states on which there is no uncertainty, that is, for which there exists a measurement that yields a particular outcome with certainty.

So how do we provide a constructive witness for the fact that the state of two systems does not `disconnect' in two separate one-system  states?  Simply by explicitly introducing a special two system state which is obtained by connecting its two outputs with a cup-shaped wire:
\[
\raisebox{-5mm}{\epsfig{figure=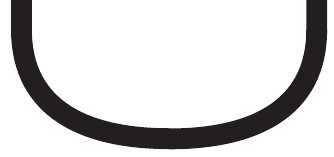,width=30pt}} 
\]
Sticking to our topological paradigm, such a cup-shaped state for example obeys:
\begin{equation}\label{eq:yank}
\raisebox{-6.5mm}{\mbox{\epsfig{figure=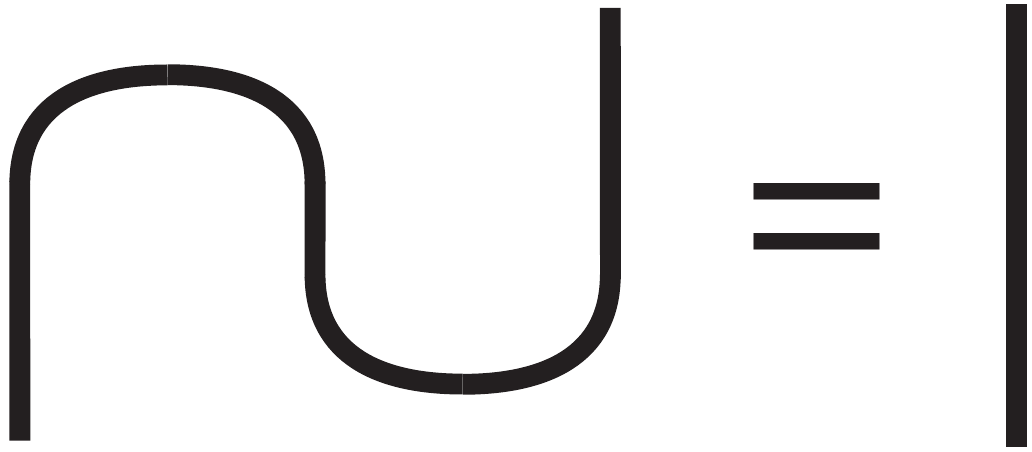,width=90pt}}}
\end{equation}
The equivalent symbolic expression for this equation would be: 
\[
(\raisebox{-0.2mm}{\epsfig{figure=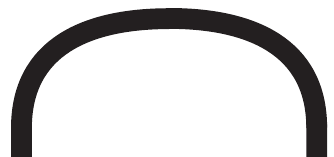,width=20pt}}\otimes 1)\circ(1\otimes \raisebox{-1.0mm}{\epsfig{figure=Cup.pdf,width=20pt}}) =1
\]
where $1$ stands for a single straight wire and $\raisebox{-0.2mm}{\epsfig{figure=Cap.pdf,width=20pt}}$ is obtained simply by flipping $\raisebox{-1.0mm}{\epsfig{figure=Cup.pdf,width=20pt}}$ i.e.~its adjoint.   We obtain a strict \em dagger-compact category \em \cite{AC1,AC2}.

\subsection{Deriving physical phenomena}

We assumed the existence of an adjoint for any box and represent it via flipping.  Cup- and cap-shaped wires also enable us to `define' the \em transpose \em which we depict by rotating a box 180${}^o$:
\begin{center}
\epsfig{figure=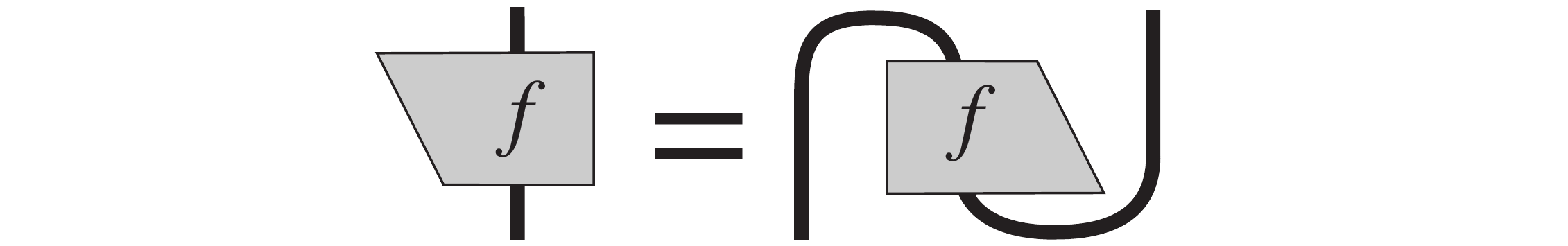,width=168pt}
\end{center}
It then immediately follows that we have:
\begin{center}
\epsfig{figure=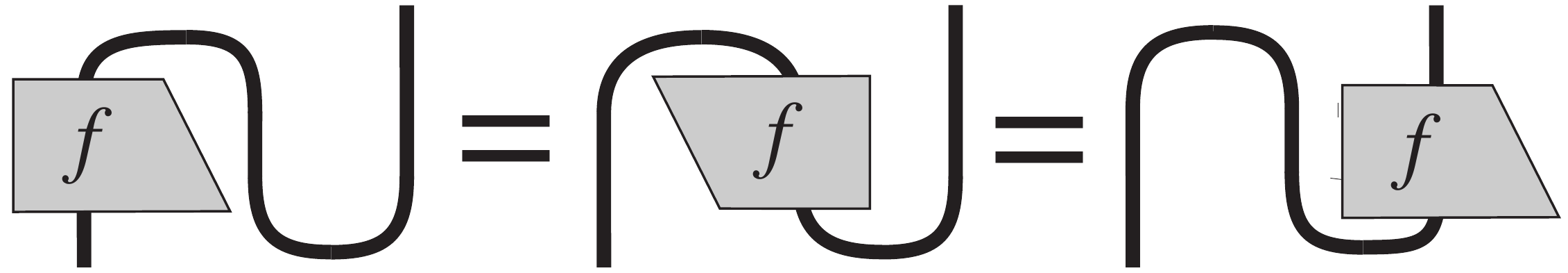,width=168pt}
\end{center}
that is, we can slide boxes across cup- and cap-shaped wires. Going berserk, 
\begin{center}
\epsfig{figure=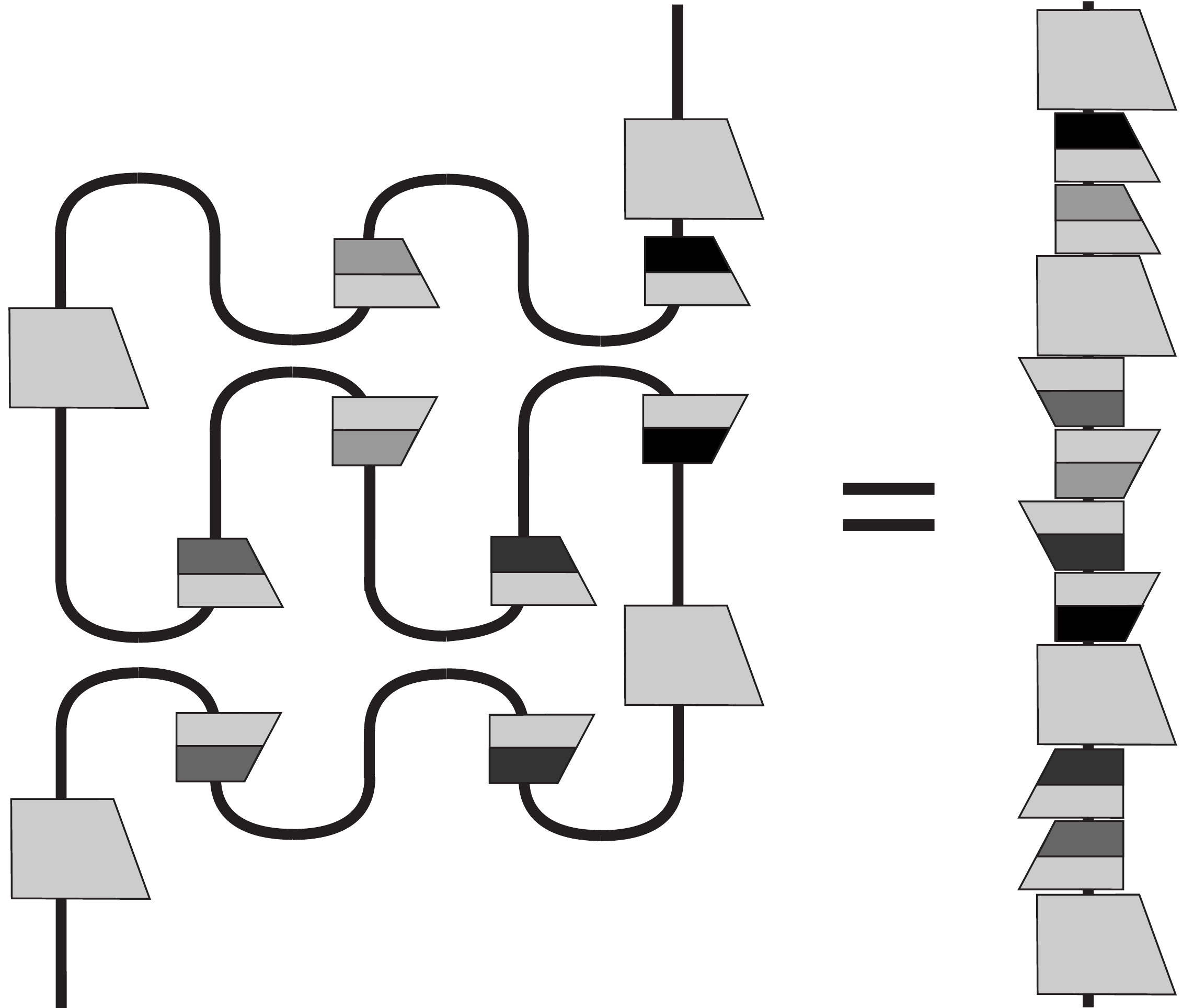,width=180pt}
\end{center}
that is, we can treat the entire graphical calculus for dagger compact categories in terms of beads which slide on wires. Now for some physics. We have:
\begin{center}
\epsfig{figure=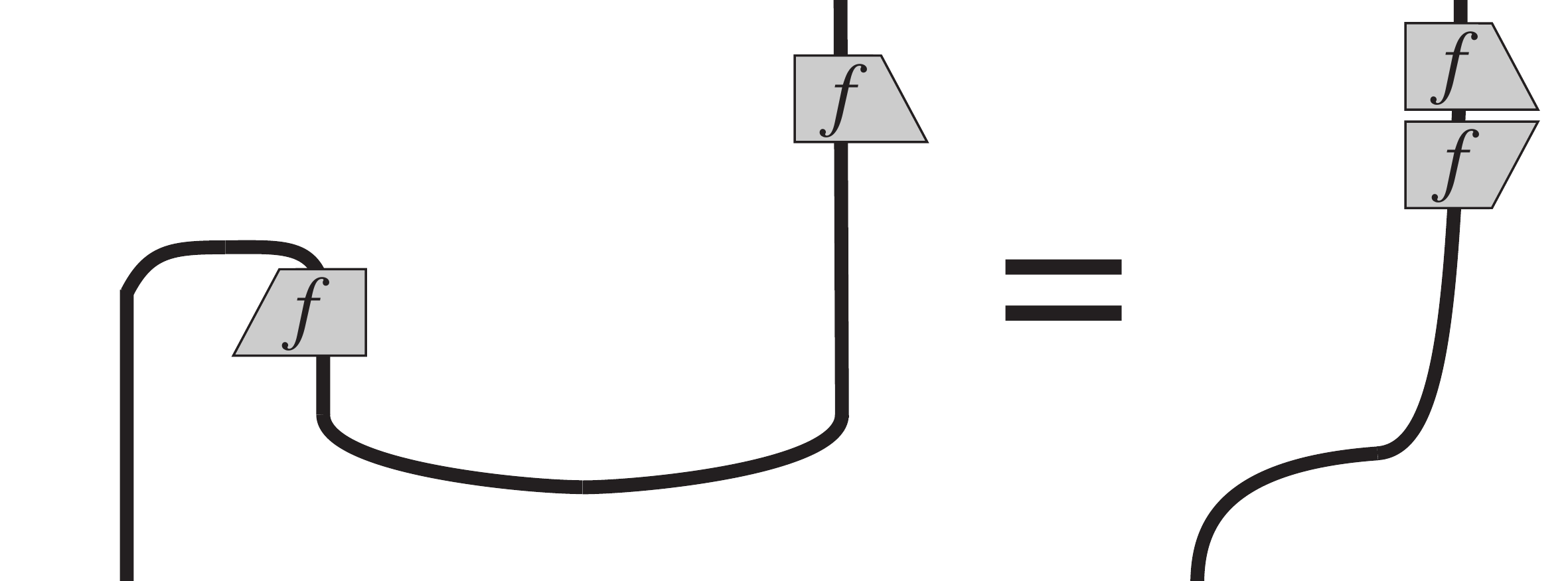,width=192pt}
\end{center}
and we choose $f$ such that its composite with its adjoint yields the identity, something to which we refer as \em unitarity\em. Hence:
\begin{center}
\epsfig{figure=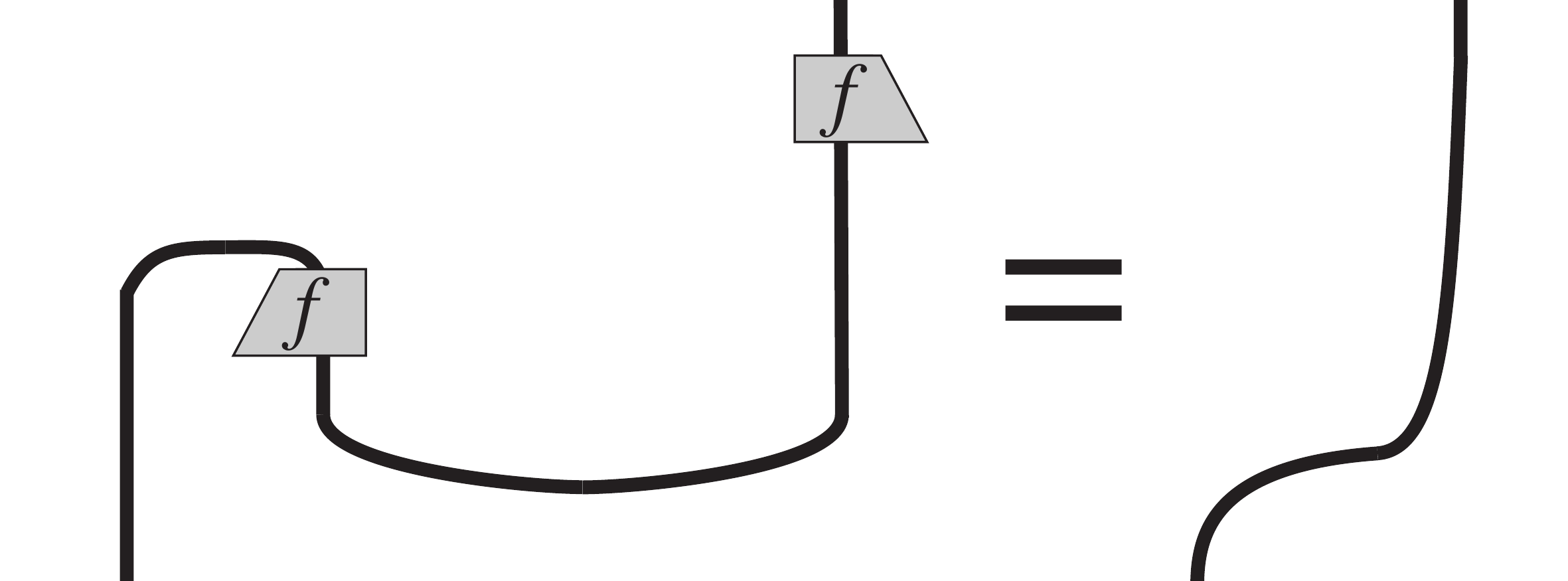,width=192pt}
\end{center}
Introducing agents Alice and Bob yields \em quantum teleportation\em: 
\begin{center}
\epsfig{figure=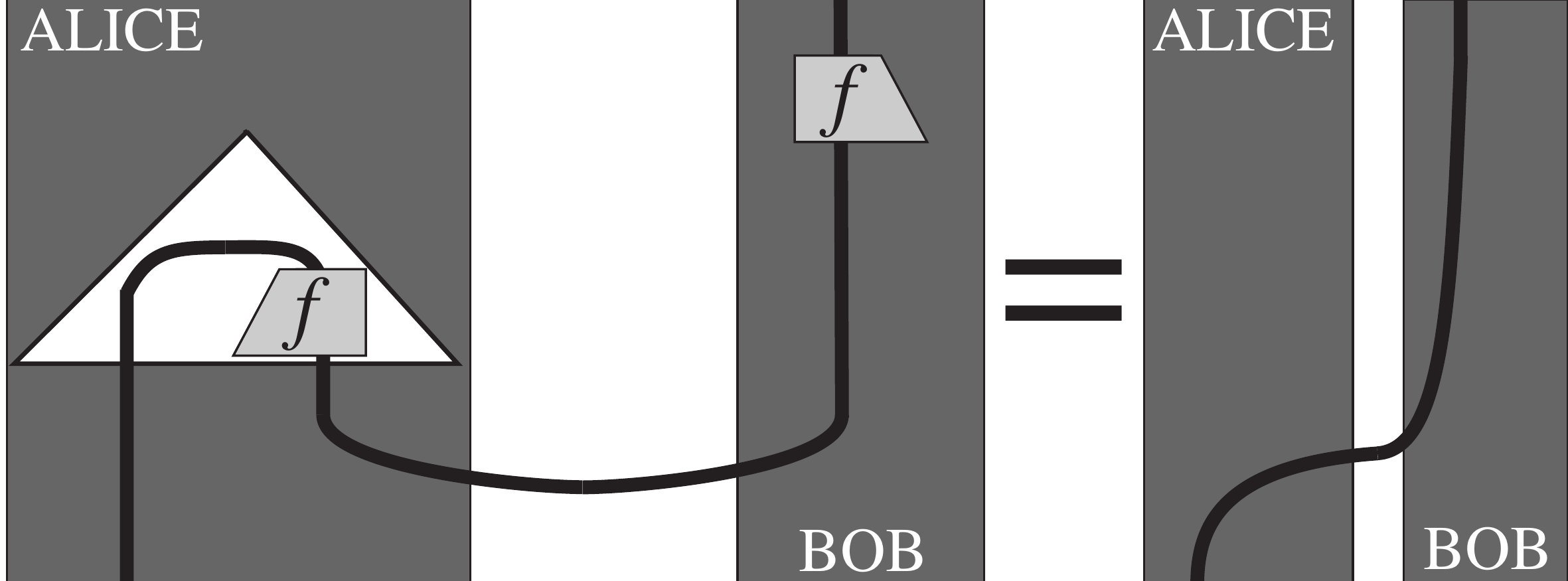,width=192pt}  
\end{center}

Note that, given that the quantum mechanical formalism was born in 1932, that this phenomenon took 60 years to be discovered \cite{Tele}.  The standard quantum mechanical formalism provides no indication whatsoever that something like this would be possible, so one had to rely on sheer luck to discover it.

A more detailed discussion of this graphical derivation and its physical interpretation is in \cite{Kindergarten,ContPhys}.  Similarly we derive another quantum mechanical feature,  the \em entanglement swapping \em protocol \cite{Swap}:
\begin{center}
\epsfig{figure=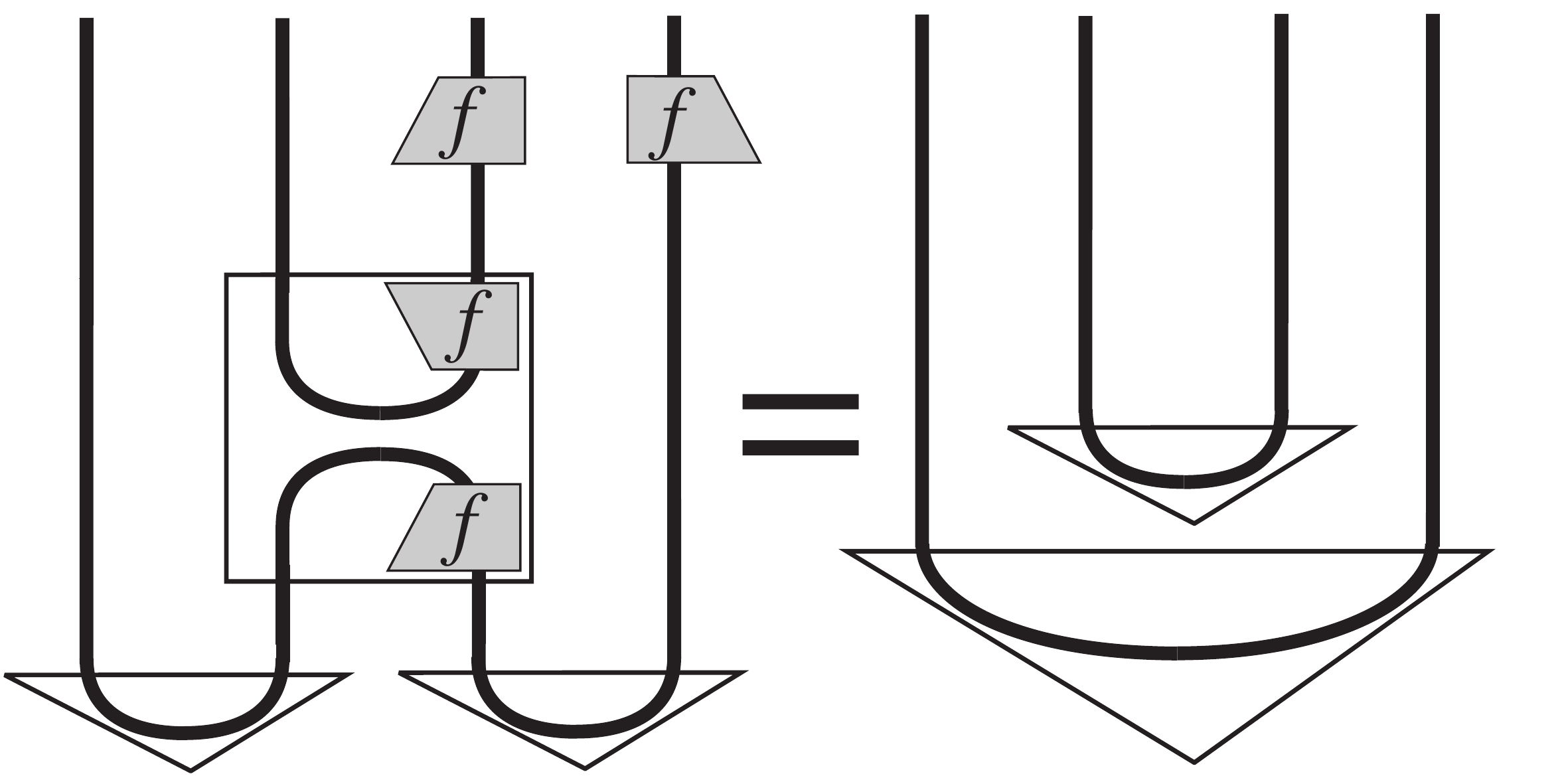,width=192pt}
\end{center}
So how much quantum mechanics can we derive in this calculus? 
 
\subsection{Logical completeness wrt Hilbert spaces}\label{sec:completeness}

The diagrammatic language presented above is directly related to the symbolic notion of a dagger compact category as follows:

\begin{thm}[Kelly-Laplaza; Selinger  \cite{KellyLaplaza,SelingerCPM}]
An equational statement between expressions in the dagger compact   categorical language  holds if and only if  it is derivable in the above described graphical calculus.
\end{thm} 

Evidently there are many dagger compact categories, to mention two:
\bit
\item Wires represent finite dimensional Hilbert spaces, boxes linear maps, the dagger is the linear algebraic adjoint, sequential composition is ordinary function composition, and the parallel composition of wires is the tensor product while parallel composition of boxes is the Kronecker product.
\item Wires represent sets, boxes relations, the dagger is the relational converse, sequential composition is composition of relations, and parallel composition is the cartesian product.
\eit
The description of the compact structure for each of these as well as some more examples can be found in \cite{CatsII}.  Evidently these two examples have very different spaces and one would evidently not associate sets and relations with quantum processes.  Hence one could could wonder how much one can actually derive in (the graphical calculus for) dagger compact categories.  The answer is surprising.
 
\begin{thm}[Hasegawa-Hofmann-Plotkin; Selinger  \cite{HasegawaHofmannPlotkin,SelingerCompleteness}] 
An equational statement between expressions in  dagger compact    categorical language  holds if and only if  it is derivable in the  dagger compact category of finite dimensional Hilbert spaces, linear maps, tensor product  and linear algebraic adjoints.
\end{thm}

To put this in more quantum physics related terms, any equation involving:
\bit
\item  states, operations, effects, ...
\item Bell-state, Bell-effect, transposition, conjugation, ...
\item inner-product, linear-algebraic trace, Hilbert-Schmidt norm, ... 
\item adjoints (e.g.~self-adjointness and unitarity), projections, positivity, ...
\item complete positivity (cf.~\cite{SelingerCPM}), ...
\eit
holds  in quantum theory if and only if it can be derived in the graphical language.  

\section{Natural language process logic}\label{sec:compdistmeaning}

Before continuing with the further development of quantum process logic, we turn our attention on something completely different: meaning in natural language, in particular, the from-word-meaning-to-sentence-meaning process.  Meaning here  manifestly goes beyond simply assigning truth values to sentences.  

\subsection{From word meaning to sentence meaning}\label{sec:wordtosent}

Consider as given the meanings of words.  This can mean many things, for example, one has a dictionary available.  On the other hand, there are no dictionaries for entire sentences.  So how de we know what a sentence means?  There must be some kind of mechanism, used by all of us,  for transforming the meaning of words into the meaning of a sentence, since surely, we all understand sentences that we may have never heard before in our lives, provided we understand all of its words.  

There is a technological side to this.  Search engines such as google and other natural language processing tools also have an understanding of meanings of words which they use to provide us with the most relevant outputs for our queries.
The model of word meaning which these engines employ enables them to produce outputs that include words that are closely related to the words in our query, i.e.~there doesn't have to be an exact match. 

However, searching on Google for ``I want something that allows me to go faster than when I only use my legs'' returns among its top hits: ``Difference Between Oxycontin and Oxycodone'', ``What are good ways for a girl to [XXX]'', ``How to Sprint Faster: 6 steps - wikiHow', ``My Story - Onelegtim.com - Retired Police Officer \& [...]'' and ``Golf Swing Power: What Your Legs Should Be Doing [...]''. Neither of these point me in the direction of appropriate vehicles that would serve my purpose, so clearly there is no understanding of the meaning of my query. The reason is the lack of a theory that produces the meaning of a sentence from the meanings of its words, whatever the manner  is in which we describe the meaning of words.

Now, representing grammatical types of words by wires and their meanings by state-boxes we can depict a string of words as:
\begin{center}
\epsfig{figure=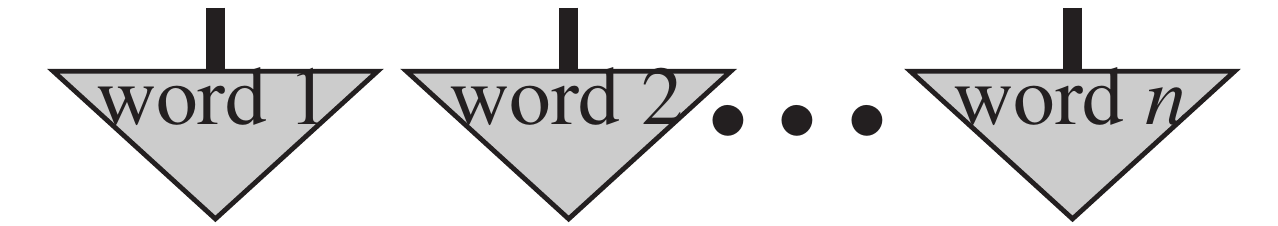,width=120pt}     
\end{center}
But the overall type, i.e.~the overall wire structure, depends on the grammatical structure of the sentence.   However, sentences with different grammatical structure may have the same meaning, and more general, we would like to have a fixed type for the meaning of all sentences.  Hence there is some process, the from-word-meaning-to-sentence-meaning process, which transforms the meanings of the string of words in the meaning of the sentence made up from these:
\begin{center}
\epsfig{figure=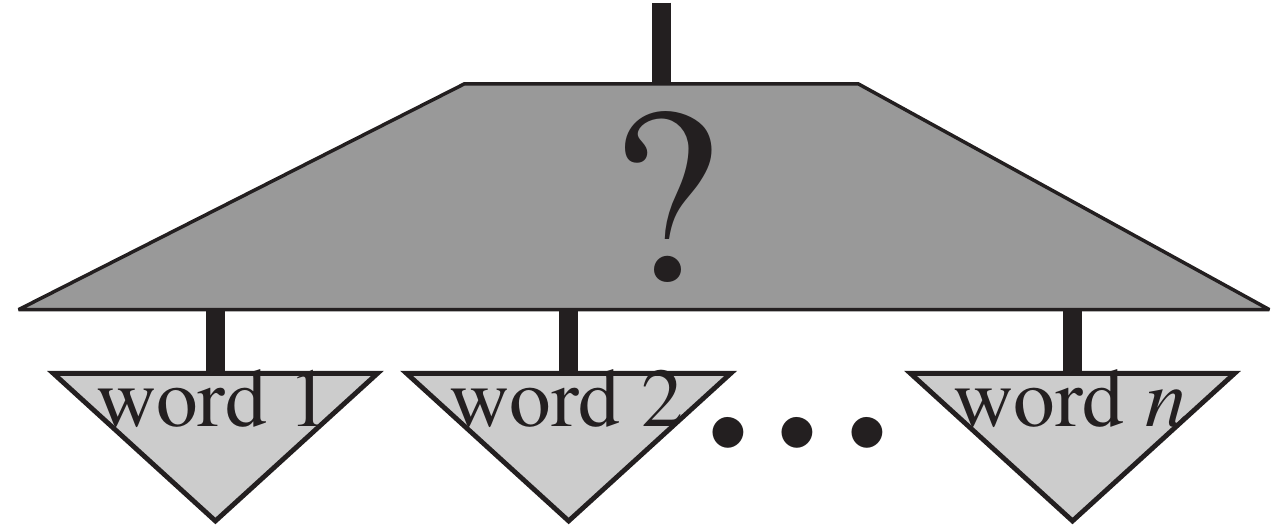,width=120pt}
\end{center}
What drives this process?  That is, given a string of words, what mediates their interaction?  The answer is obvious:
\begin{center}
\epsfig{figure=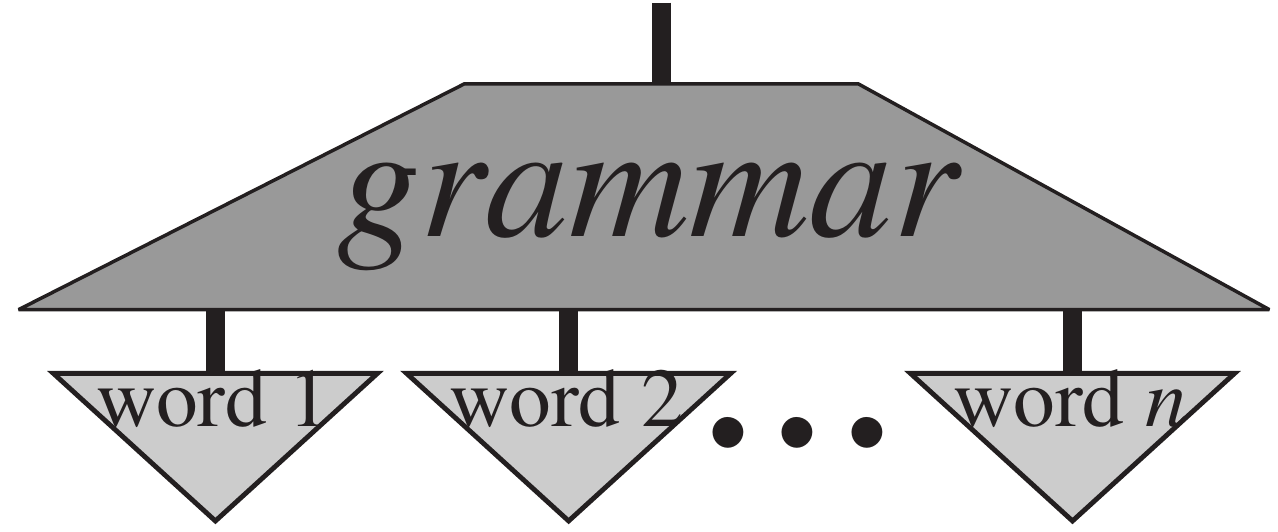,width=120pt}   
\end{center}
since grammatically incorrect sentences have no clear meaning anyway.

We can now describe the problem for from-word-meaning-to-sentence-meaning processes in more precise terms:
\bit
\item Given a theory of word meaning, and given a theory of grammar, how can we combine these into an algorithm which produces the meaning of sentences from the meanings  of its words?
\eit
As already mentioned, this problem was addressed by Clark, Sadrzadeh and the author in \cite{CCS,CSC}.  Let's stay at an abstract level a bit longer, before we will describe concrete theories of word meaning and grammar.  What is a verb?  A transitive verb is something that requires an object and a subject in order to yield a  grammatically correct sentence.  So we can think of a transitive verb as a process with three wires, two  respectively requiring an object and a subject, and one producing the  sentence:
\begin{center}
\epsfig{figure=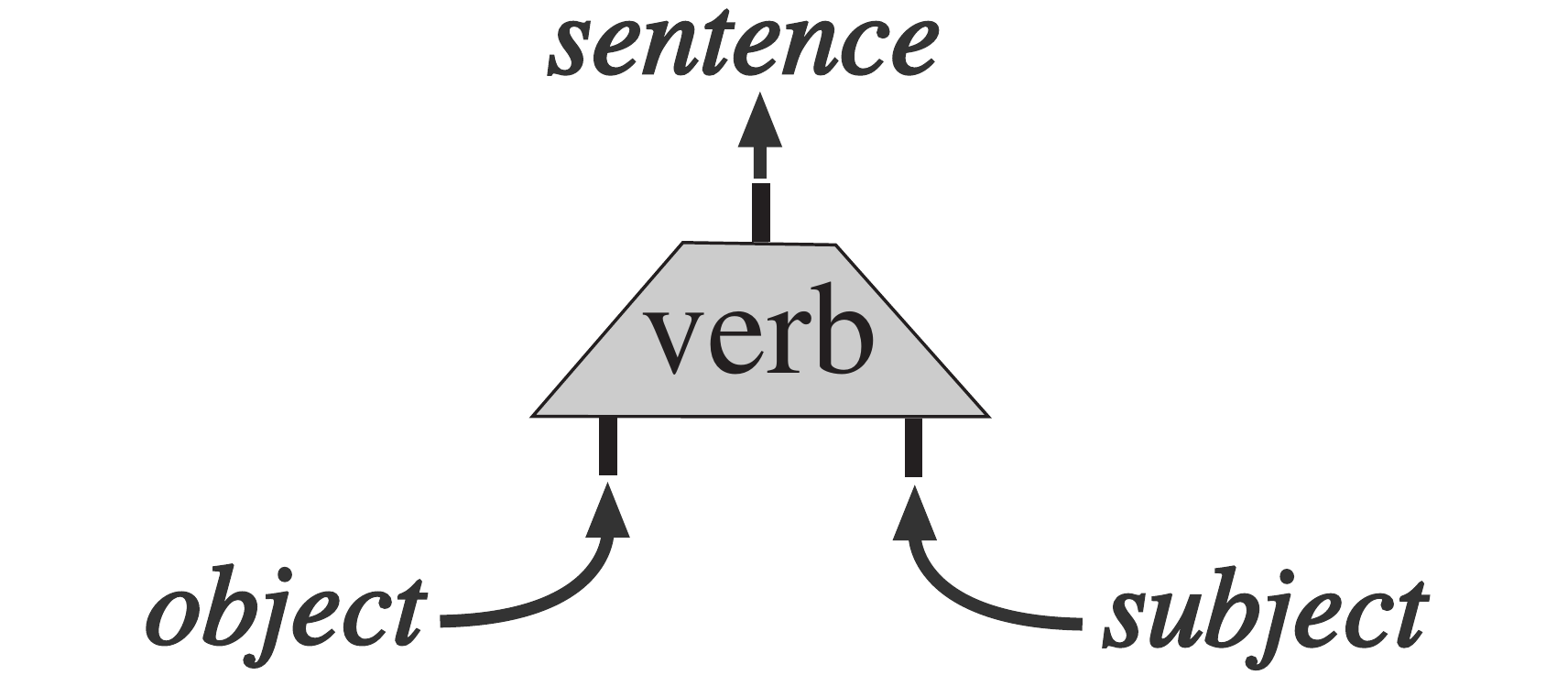,width=140pt}    
\end{center}
Since we rather represent a verb as a state we can use transposition, as defined above, to turn inputs into outputs and represent the verb as:
\begin{center}
\epsfig{figure=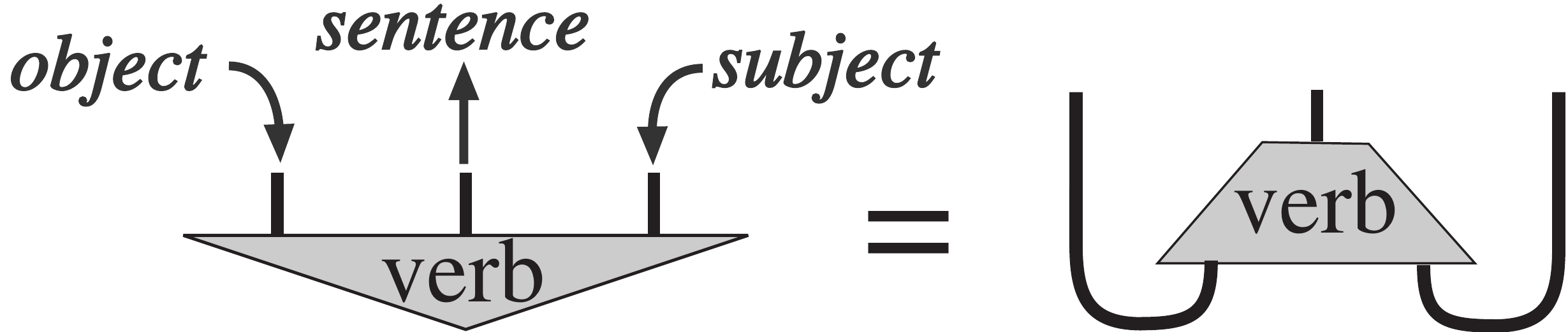,width=200pt}   
\end{center}
You may ask where these  cups suddenly come from, but here we already anticipate the description of grammatical structure that we discuss below.  

Note in particular also that for these kinds of word-states we again have:
\begin{center}
\epsfig{figure=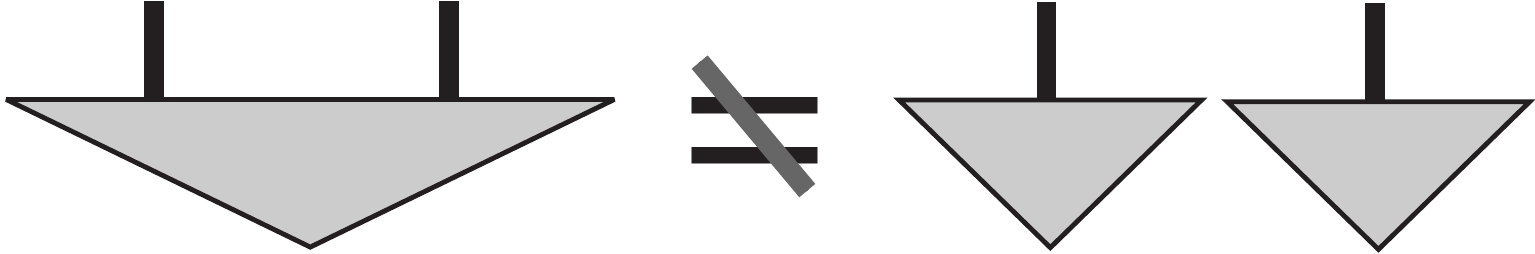,width=130pt}
\end{center}
since otherwise, for the case of a transitive verb, the meaning of the sentence would not depend on the meanings of the nouns, which could have dramatic consequences.  For example, considering the verb `hate', it would be sufficient for one person to hate another person in order for everyone to hate everyone.

\subsubsection{A theory for word meaning}

The current dominant theory of word meaning for natural language processing tasks is the so-called \em distributional \em or \em vector space model \em of meaning \cite{Schuetze}.  It takes inspiration from WittgensteinÕs  philosophy of Ômeaning is useÕ \cite{Wittgenstein}, whereby meanings of words can be determined from their context,  and works as follows.  One fixes a collection of $n$  words, the \em context words\em, and considers an $n$ dimensional vector space with chosen basis where each basis vector  represents one of the context words.  Then one selects a huge body of written text, the \em corpus\em. E.g.~the internet, all editions of a certain newspaper, all novels, the British National Corpus\footnote{This can be accesses at http://www.natcorp.ox.ac.uk/}  which is a 100 million word collection of samples of written and spoken language from a wide range of sources, etc.  Next one decides on a \em scope\em, that is, a small integer $k$, and for each context word $x$ one counts how many times $N_x(a)$ a word $a$ to which one wants assign a meaning occurs at a distance of at most $k$ words from $x$.  One obtains a vector $(N_1(a), \ldots, N_n(a))$, which one normalizes in order to obtain $(\phi_1(a), \ldots, \phi_n(a))$, the \em meaning vector \em of $a$.  Now, in order to compare meanings of words, in particular, how closely their meanings are related, one can simply compute the inner-product of their meaning vectors.  

\subsubsection{A theory of grammar}

Algebraic gadgets that govern grammatical types have been around for quite a bit longer \cite{Ajdukiewicz,Bar-Hillel,Chomsky, Lambek0}.  There are several variants available, each with their pro's and con's; here we will focus on Lambek's \em pregroups \em \cite{Lambek1}.   Philosophically, these algebraic gadgets trace back to FregeÕs principle that the meaning of a sentence is a function of the meaning of its parts \cite{Frege}.  However, this is only manifest in that these algebras all have a composition operation that allows to build larger strings of words from smaller strings of words.  These algebras also  have a relation $\leq$ where $a\cdot \ldots \cdot z\leq t$ means that the string of types $a \ldots  z$ has as its overall type $t$.  For example, $n \cdot tv \cdot n\leq s$ expresses the fact that a noun, a transitive verb and a noun make up a sentence $s$.  Finally, there are additional operations subject to certain laws which make up the actual structure of the algebra, and these would allow one to derive correct statements such as $n \cdot tv \cdot n\leq s$.

For the specific case of pregroups, these additional operations are a \em left inverse \em ${}^{-1}(-)$ and a \em right inverse \em $(-)^{-1}$, subject to $x\cdot {}^{-1}(x)\leq 1$ and $(x){}^{-1}\cdot x\leq 1$ where $1$ is the unit for the composition operation, as well as to $1 \leq {}^{-1}(x) \cdot x$ and $1\leq  x\cdot (x){}^{-1}$.   Now we have to assign grammatical types to the elements of a pregroup.  Some will be atomic, i.e.~indecomposable, while others like transitive verbs will be assigned compound types.  Concretely, $tv = {}^{-1}(n) \cdot s \cdot (n)^{-1}$,  hence
\[
n\cdot tv \cdot n =  n\cdot \left({}^{-1}(n) \cdot s \cdot (n)^{-1}\right) \cdot n=  \left(n\cdot {}^{-1}(n)\right) \cdot s \cdot \left((n)^{-1} \cdot n\right) \leq 1\cdot s\cdot 1 =s\,,
\]
so the string of types `noun transitive verb noun' indeed makes up a grammatically correct sentence.  We can depict this computation graphically as follows.  We start with five systems of respective types $n$, ${}^{-1}(n)$,  $s$, $(n)^{-1}$ and $n$:
\begin{center}
\epsfig{figure=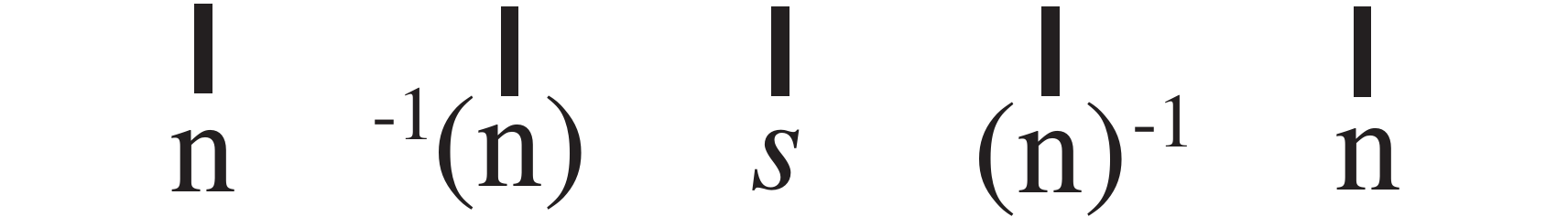,width=140pt}
\end{center}
Then, we use caps to indicate that $n$ and ${}^{-1}(n)$, and, $(n)^{-1}$ and $n$, cancel out:
\begin{center}
\epsfig{figure=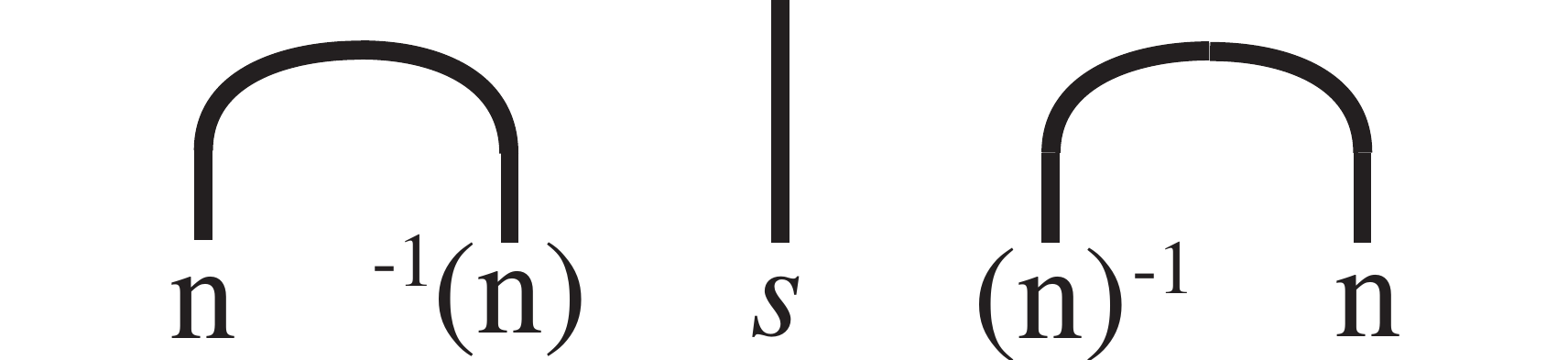,width=140pt}
\end{center}
so that at the end the only remaining system is the sentence type.  The caps here represent the equations $x\cdot {}^{-1}(x)\leq 1$ and $(x){}^{-1}\cdot x\leq 1$.  In fact, this is not just an analogy with the graphical language of  compact categories. Pregroups \em are \em in fact  compact categories!  To see this, any partial order is a category, the composition provides the tensor, and while equations $x\cdot {}^{-1}(x)\leq 1$ and $(x){}^{-1}\cdot x\leq 1$ provide caps, equations $1 \leq {}^{-1}(x) \cdot x$ and $1\leq  x\cdot (x){}^{-1}$ provide cups.  More details on this are in \cite{CSC}.  The reason that there are two kinds of caps and cups is the fact that we are not allowed to change the order of words in a sentence while  two physical systems do not come with some ordering.  In category-theoretic terms, here we are dealing with a \em non-symmetric tensor\em. 

\subsection{Combining theories}

The structural similarity between the pregroup theory of grammar, and the vector spaces for word meaning when organized as a dagger compact category, is exactly what we will exploit to explicitly construct the  from-word-meaning-to-sentence-meaning process.  We consider the graphical representation of the proof of grammatical correctness of a sentence, substitute the sentence types by meaning vectors of the particular words we are interested in, and substitute the caps by the vector space caps, so we obtain:
\begin{center}
\epsfig{figure=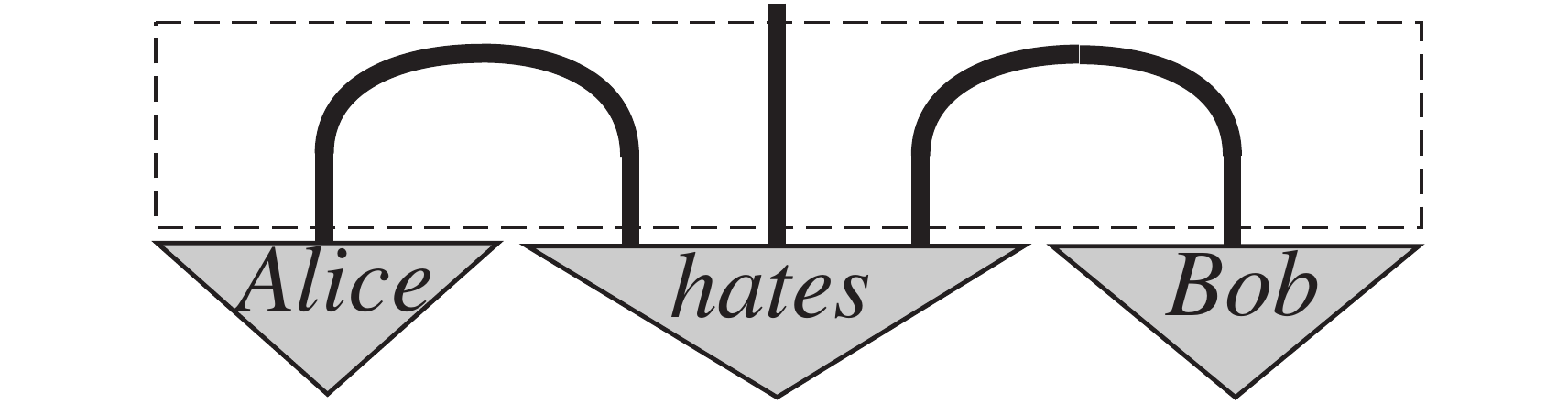,width=140pt} 
\end{center}
where the dotted line indicates the linear map that when applied to the vector $\overrightarrow{Alice}\otimes\overrightarrow{hates}\otimes\overrightarrow{Bob}$ produces the vector that we take to be the meaning of a sentence.  By rewriting this using transposition, as in Section \ref{sec:wordtosent}, the verb now acts as a fuction on the object and the subject:
\begin{center}
\epsfig{figure=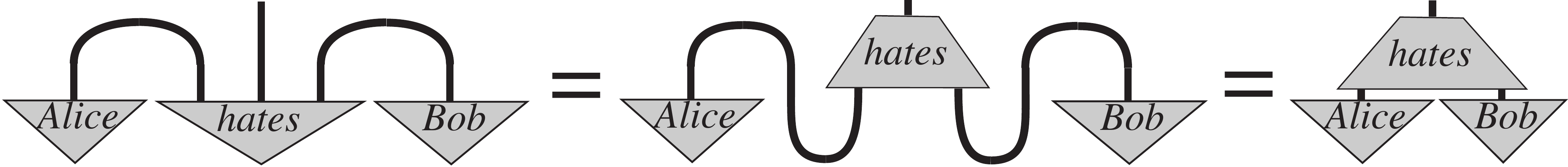,width=346pt} 
\end{center}

The meanings of all sentences live in the same vector space so we can again simply use the inner-product to measure their similarity.  Grefenstette and Sadrzadeh have recently exploited this theory for standard natural processing tasks and their method outperforms all existing ones \cite{GrefSadr}.

What about the cups?  They can be used to model `special words' like ``does'' and ``not'', which have a clear `logical' meaning.  Here is an example of this:
\begin{center}
\epsfig{figure=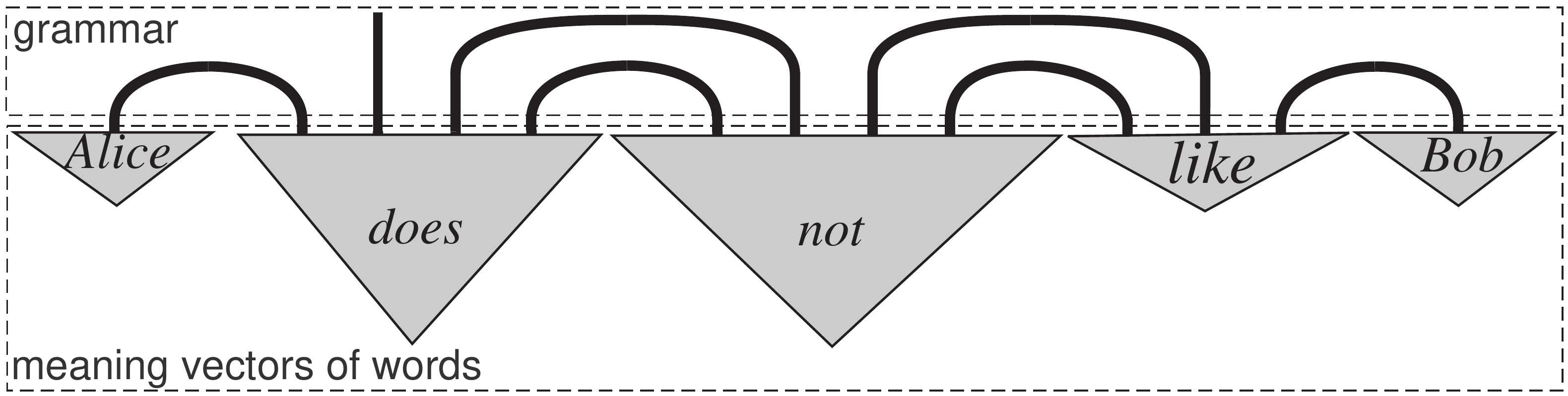,width=260pt}
\end{center}
As above, the wire structure here is obtained from the types of these words according to the pregroup grammar.  Using cups we can model the meaning of `does', that is, `does nothing really', and `not', that is, `negates meaning', for which we use an input-output $not$-box that does just that:
\begin{center}
\epsfig{figure=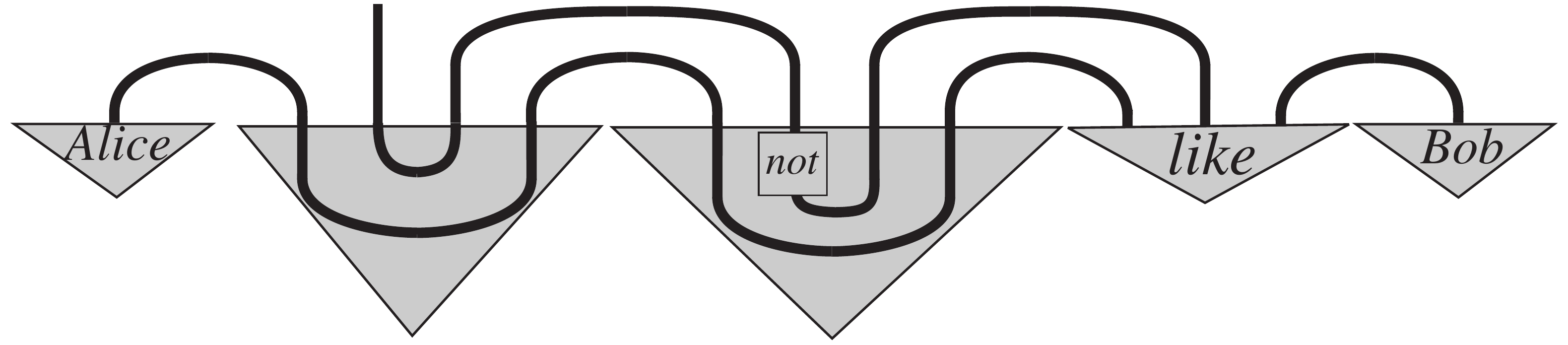,width=260pt}
\end{center}
and then we can simply use homotopy to compute:
\begin{center}
\epsfig{figure=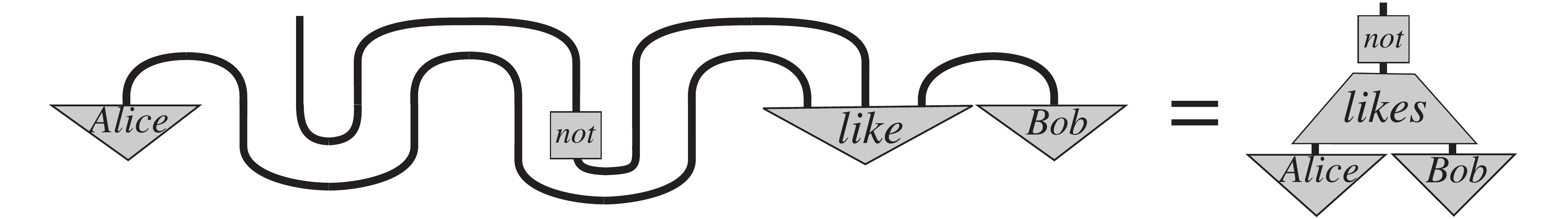,width=340pt}
\end{center}
which is exactly what we would expect the meaning of Alice doesn't liking Bob to be: the negation of Alice liking Bob.  This example also shows how the wires are mediating the `flow' of word meaning in sentences.  They allow for the words Alice and like, while far apart in the sentence, to interact.  

Turning things upside-down, one can now ask the question: why are there algebraic gadgets that describe grammatical correctness, i.e.~why do these even exist.  Our theory of word meaning explains this: they witness the manner of how word meanings interact to form the meaning of a sentence.

\subsection{An aside: quantizing grammar} 


An interesting analogy arises, which was first observed by Louis Crane, and which is discussed in detail in \cite{PrelSadr}.    An important area of contemporary mathematics is the study of Topological Quantum Field Theory (TQFT) \cite{Atiyah,BaezDolan,Baez}.  While it takes its inspiration from quantum field theory, it has become an area of research in its own right, mainly within topology.  The object of study is a monoidal functor:
\[
F: n{\bf Cob}\to {\bf FVect}_{\mathbb{K}}::
\raisebox{-6mm}{\epsfig{figure=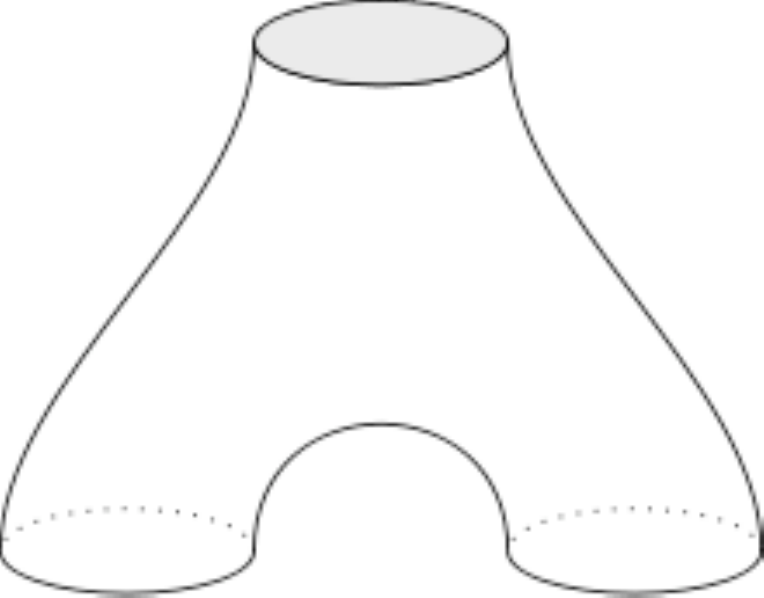,width=50pt}}\mapsto V
\]
from the compact category of closed $(n-1)$-dimensional manifolds  with diffeomorphism classes of $n$-dimensional manifolds connecting the closed $(n-1)$-dimensional manifolds as morphisms, to the compact category of vector spaces over some field $\mathbb{K}$.\footnote{If the field has a non-trivial involution then this category has a dagger too ($\not=$ transposition).} Now, rather than taking a category of topological structures as domain, we can take a pregroup as domain, i.e.~a category of grammatical structures, and obtain a \em grammatical quantum field theory\em: 
\[
F: Pregroup\to {\bf FVect}_{\mathbb{R}_+}:: 
\raisebox{-3mm}{\epsfig{figure=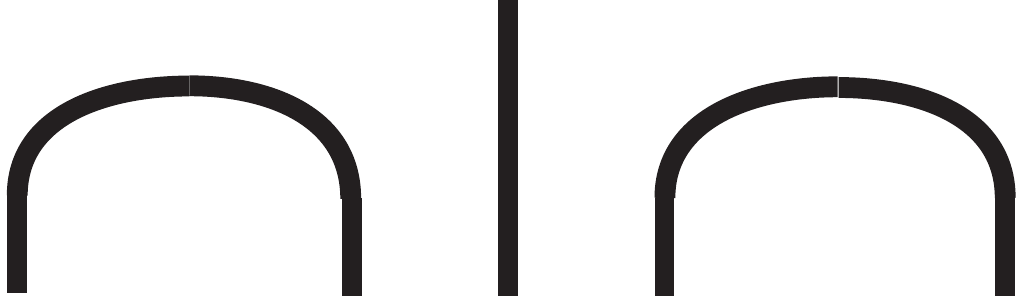,width=100pt}}\mapsto V
\]


\section{Quantum process logic - Take IIb}

Dagger compact categories  capture a substantial number of quantum mechanical concepts, and the dagger compact category ${\bf FdHilb}$ related to the von Neumann model described in Section \ref{sec:completeness}, is complete with respect to them.  But they are by no means universal with respect to quantum theory, by which we can mean two different things:
\bit
\item that they do not capture all quantum mechanical concepts, and,
\item that the language is not rich enough to describe all processes in ${\bf FdHilb}$.
\eit
Examples of concepts that are not captured by dagger compact language are the  classical data obtained in measurements, observables themselves, and relationships between these e.g.~complementarity.  Examples of ${\bf FdHilb}$-processes  not expressible in dagger compact language are basic quantum computational gates such as the CNOT-gate, phase-gates etc.  We will now present an extended graphical language which does capture all of these.  This was established in a series of papers by Pavlovic, Paquette,  Duncan, and the author in \cite{CPav, CPaqPav, CD1, CD2}.  The calculus was also rich enough to address a number of concrete quantum computational and quantum foundational problems e.g.~see \cite{DP2,CES,Clare,BoixoHeunen,CDKZ}.

Rather than only allowing for wires we allow for `dots' at which wires branch into multiple wires, or none.  We refer to these dots as ...
\[ 
\mbox{\it `spiders'}\ = \left\{\raisebox{-0.44cm}{$\underbrace{\overbrace{\epsfig{figure=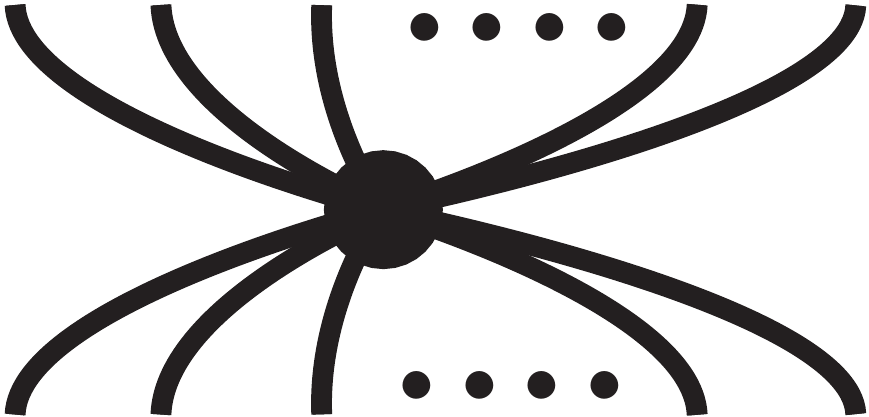,width=65pt}}^{\scriptstyle{m}}}_{\scriptstyle{n}}$}
\right\}_{n, m}
\]
So what is the analogue the topological calculus with cups and caps, and in particular, eq.(\ref{eq:yank})? Similarly to `however one bends a wire, it  still remains just a wire that acts as an identity', any web of spiders with the same overall number of inputs and outputs, independent of how the web is build up, is again the same.  So  for any $k>0$:
\[
    \underbrace{\overbrace{\epsfig{figure=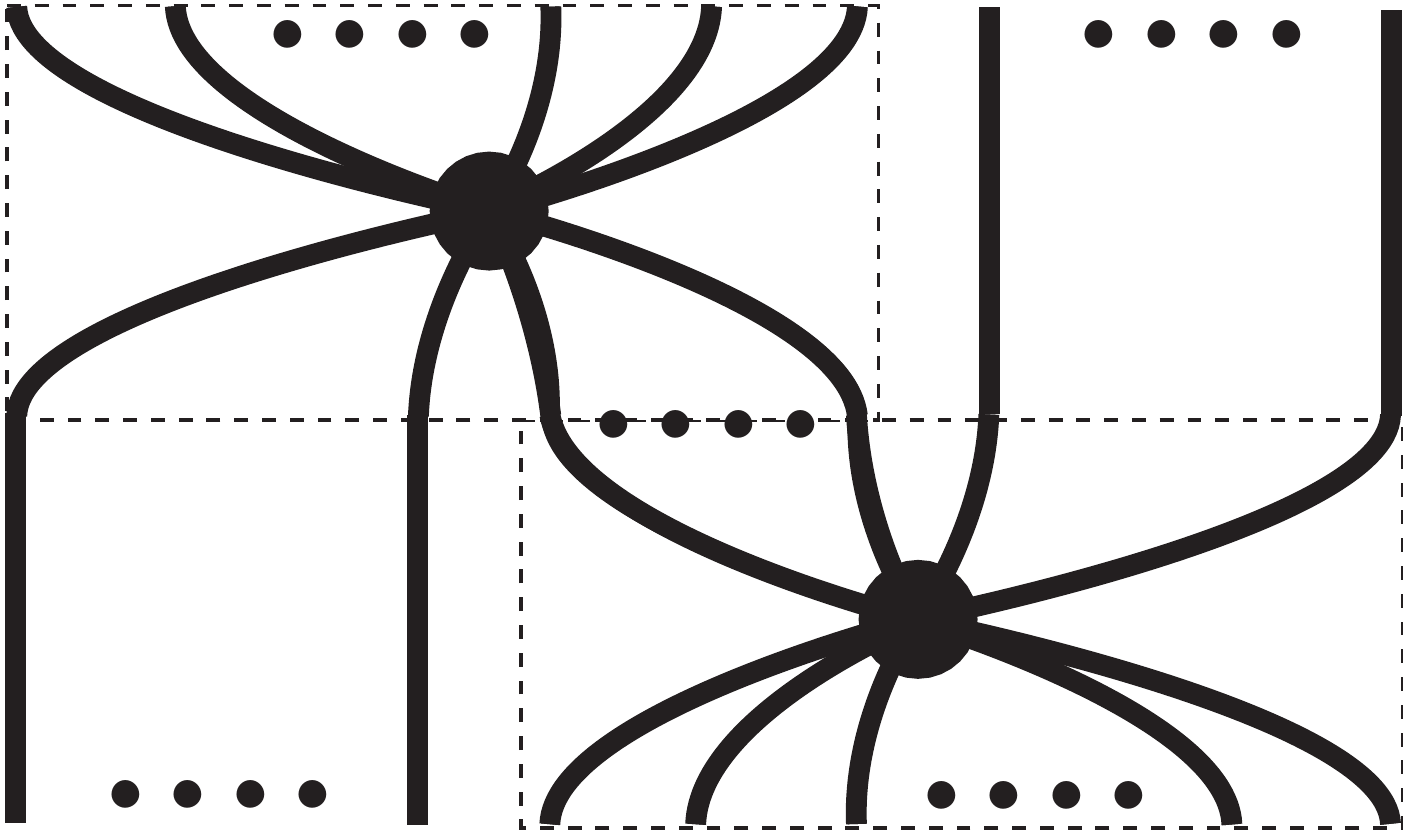,height=65pt}}^{\scriptstyle{m+m'-k}}}_{\scriptstyle{n+n'-k}}
   \ \  \epsfig{figure=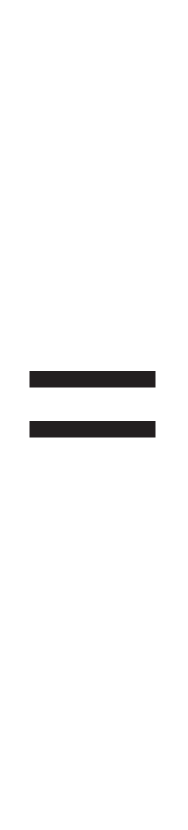,height=65pt}\ \ 
    \epsfig{figure=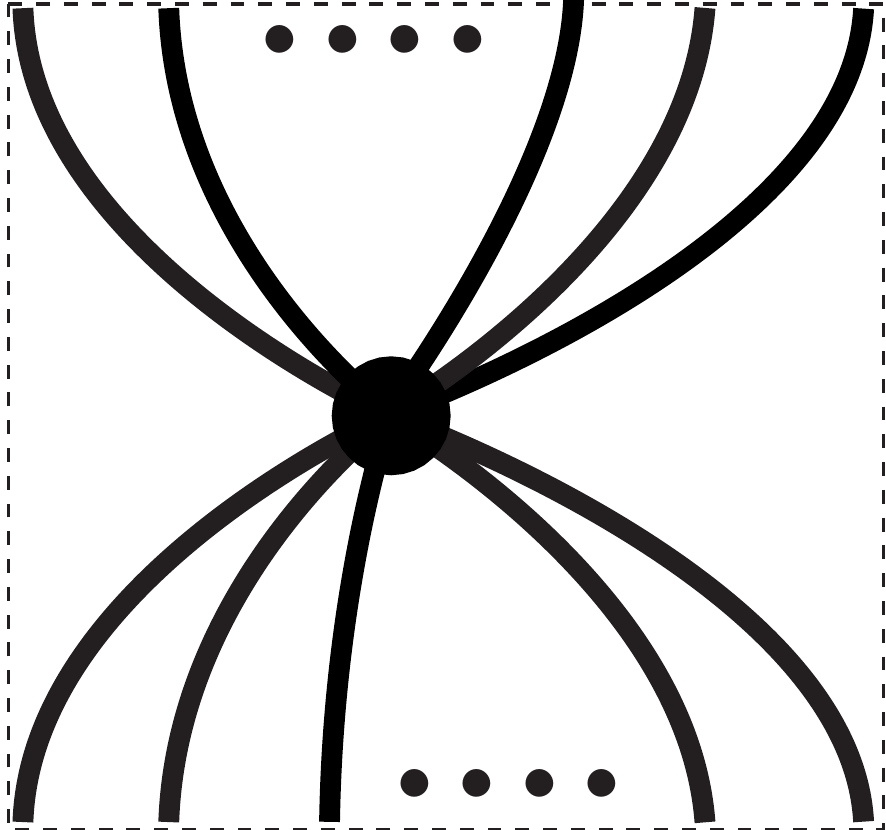,height=65pt}
\]
Hence, the rule governing spider calculus is that if two spiders `shake legs', they fuse together.  Again in other words,  it only matters what is connected to what, but not the manner in which this connection is realized.

This in particular implies that for the specific spiders:
\[
\underbrace{
\overbrace{
\epsfig{figure=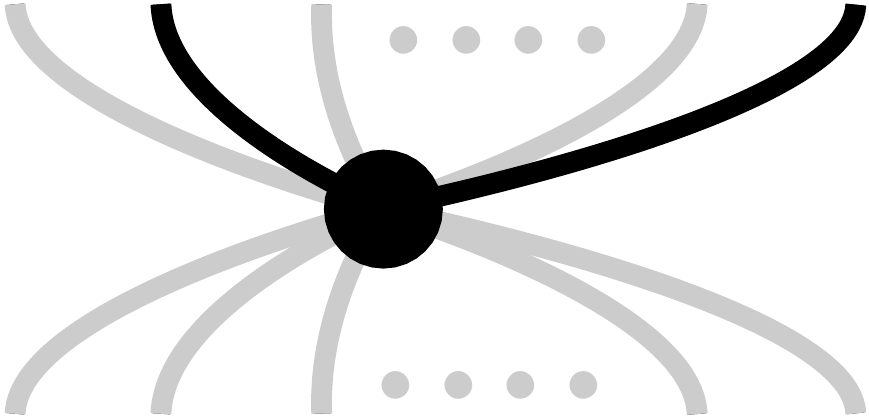,width=65pt}
}^{\scriptstyle{2}}
}_{\scriptstyle{0}}
\qquad\raisebox{5mm}{\mbox{\rm and}}\qquad
\underbrace{
\overbrace{
\epsfig{figure=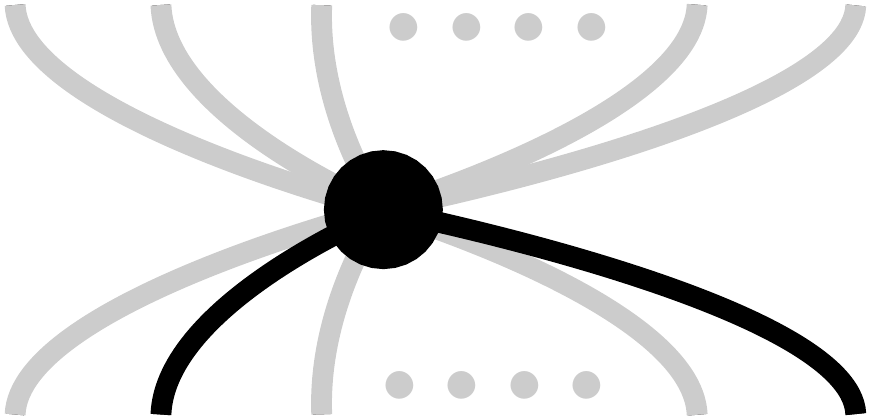,width=65pt}
}^{\scriptstyle{0}}
}_{\scriptstyle{2}}
\]
we obtain eq.(\ref{eq:yank}): 
\[
\underbrace{\overbrace{\epsfig{figure=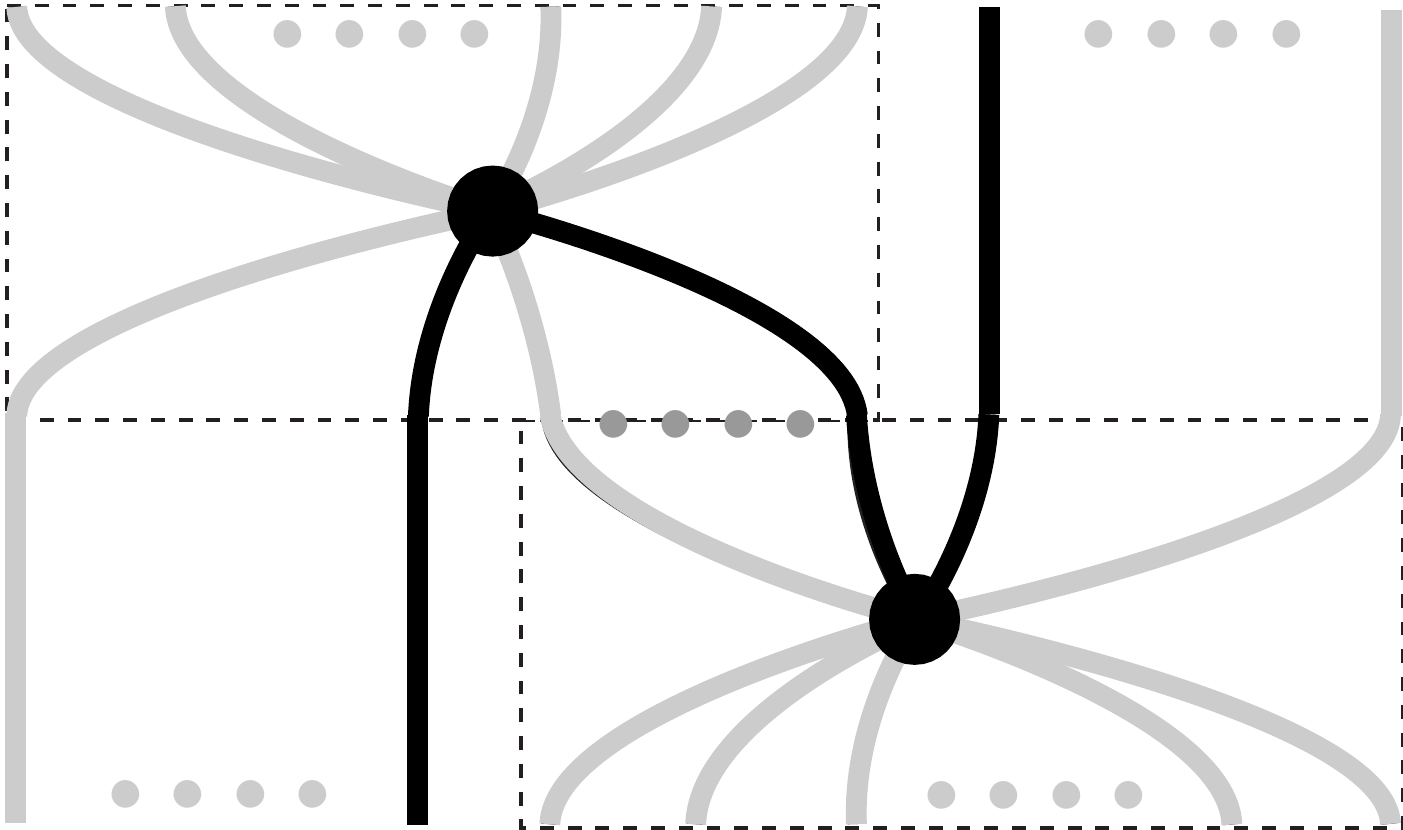,height=65pt}}^{\scriptstyle{0+2-1}}}_{\scriptstyle{2+0-1}}
   \ \  \epsfig{figure=SpiderComposition2,height=65pt}\ \ 
    \epsfig{figure=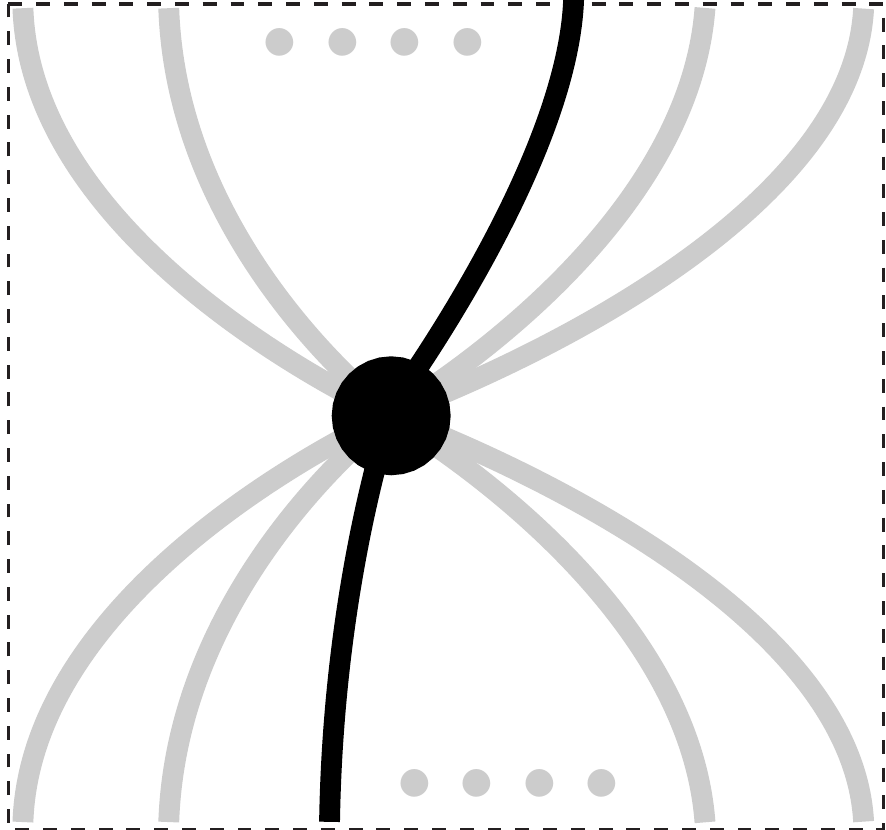,height=65pt}
\]
so \em reasoning with spiders strictly generalizes reasoning with wires\em.  

In ${\bf FdHilb}$ a family of spiders of the above kind on-the-nose captures an orthonormal  basis, which is a non-trival result.  Firstly, one can show that reasoning with those spiders is equivalent to working with a so-called dagger special commutative Frobenius algebra \cite{KockBook,Lack,CPaq}. Next one shows that these dagger special commutative Frobenius algebras in  ${\bf FdHilb}$ are the same thing as orthonormal  bases \cite{CPV}.  Since bases allow to represent observables and classical data, we almost reached our goal, except for the fact that quantum theory only becomes interesting if we consider several `incompatible' bases.

So now we consider two different families of spiders, represented by a different gray scale. What happens if a dark gray and a light gray spider which represent complementary observables `shake leggs'?  Well, their `legs fall off':
\[
\epsfig{figure=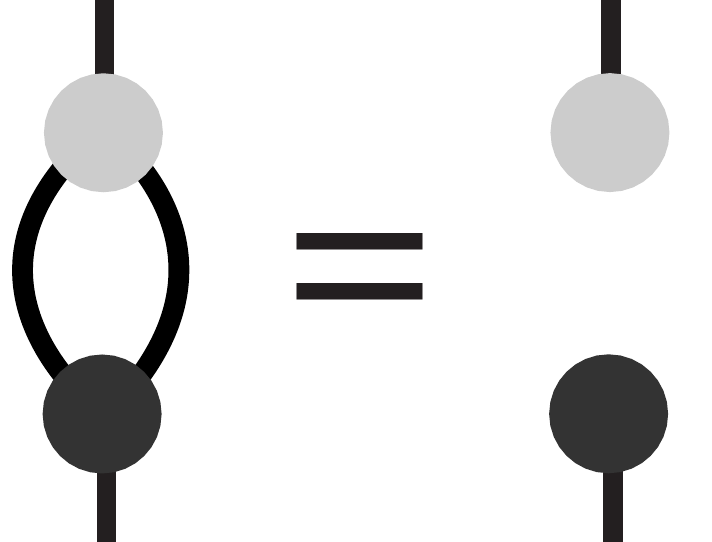,width=60pt} 
\]
This was shown by Duncan and the author in \cite{CD2}. Such a pair of differently colored spider families that interact in this manner forms the basis of a rich  calculus with many more extra features than the ones described here.  We refer the interested reader to \cite{ContPhys,CD2,CDKZ} for more details and concrete applications.

\section{The remaining challenge}


In this paper we pushed forward the idea that the diagrammatic languages describing quantum phenomena as well as meaning-related linguistic phenomena may constitute some new kind of quantitative logic.  The same logic also governs Bayesian inference, Bayesian inversion boiling down to nothing but transposition for appropriately chose cups and caps:
\[
\epsfig{figure=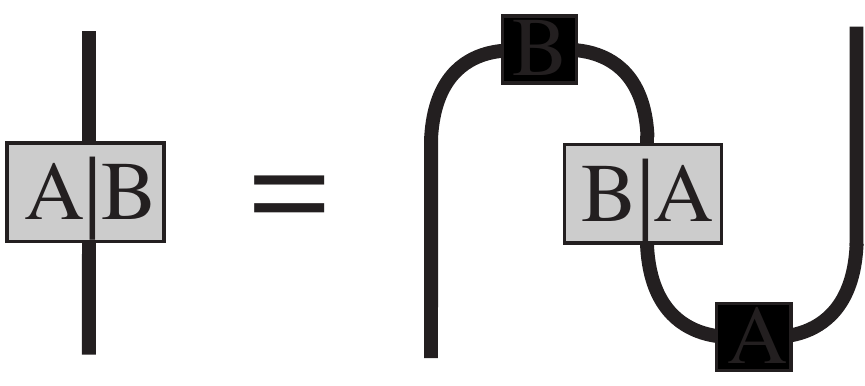,width=80pt} 
\]
This was established by Spekkens and the author in \cite{CSBayes}, to which we refer for details.
So where does traditional logic fit into this picture?

One perspective is to start with standard categorical logic \cite{LambekScott,AbrLNP,BaezLNP}.  The compact structure can then be seen as a resource sensitive variant (as in Linear Logic \cite{Girard,Seely}) which is degenerate in the sense that conjunction and disjunction coincide  \cite{RossThesis, AAAI}.  We do not subscribe (anymore) to conceiving the diagrammatic  logic as a `degenerate hyper-deductive variant' of standard logic in categorical form since this does not recognize the quantitative nor the process content.   

Rather, we would like to conceive the quantitative diagrammatic logic as `the default thing' from which traditional qualitative logic arises via some kind of structural collapse.  There are several results that could be taken as a starting point in this direction, for example, the generalization in \cite{CPaqPav} of Carboni and Walters' axiomatization of the category of relations \cite{CarboniWalters}. But since this still belongs to the world of speculation, we leave this to future writings. 

\bibliographystyle{plain}
\bibliography{ASL_paper_bibhack} 

\begin{thebibliography}{10}

\bibitem{AC1}
S.~Abramsky and B.~Coecke.
\newblock A categorical semantics of quantum protocols.
\newblock In {\em LICS Proceedings}, pages 415--425. IEEE Computer Society,
  2004.
\newblock Extended version: {a}rXiv:quant-ph/0402130.

\bibitem{AC2}
S.~Abramsky and B.~Coecke.
\newblock {Abstract physical traces}.
\newblock {\em Theory and Applications of Categories}, 14(6):111--124, 2005.

\bibitem{AbrLNP}
S.~Abramsky and N.~Tzevelekos.
\newblock Introduction to categories and categorical logic.
\newblock In B.~Coecke, editor, {\em New Structures for Physics}, Lecture Notes
  in Physics, pages 3--94. Springer, 2011.

\bibitem{Ajdukiewicz}
K.~Ajdukiewicz.
\newblock Die syntaktische konnexit\"at.
\newblock {\em Studia Philosophica}, (1):1--27, 1937.

\bibitem{NewScientist}
J.~Aron.
\newblock Quantum linguistics leap forward for artificial intelligence.
\newblock {\em New Scientist}, 208(2790):10--11, 2010.

\bibitem{Atiyah}
M.~Atiyah.
\newblock Topological quantum field theories.
\newblock {\em Publications Math{\'e}matiques de l'IH{\'E}S}, 68(1):175--186,
  1988.

\bibitem{Baez}
J.C. Baez.
\newblock {Quantum quandaries: a category-theoretic perspective}.
\newblock In D.~Rickles, S.~French, and J.T. Saatsi, editors, {\em The
  Structural Foundations of Quantum Gravity}, pages 240--266. Oxford University
  Press, 2006.
\newblock {arXiv:quant-ph/0404040}.

\bibitem{BaezDolan}
J.C. Baez and J.~Dolan.
\newblock Higher-dimensional algebra and topological quantum field theory.
\newblock {\em Journal of Mathematical Physics}, 36:6073, 1995.
\newblock {arXiv:q-alg/9503002}.

\bibitem{BaezLNP}
J.C. Baez and M.~Stay.
\newblock Physics, topology, logic and computation: a {R}osetta stone.
\newblock In B.~Coecke, editor, {\em New Structures for Physics}, Lecture Notes
  in Physics, pages 95--172. Springer, 2011.

\bibitem{Bar-Hillel}
Y.~Bar-Hillel.
\newblock A quasiarithmetical notation for syntactic description.
\newblock {\em Language}, (29):47--58, 1953.

\bibitem{Benabou}
J.~Benabou.
\newblock Categories avec multiplication.
\newblock {\em Comptes Rendus des S\'eances de l'Acad\'emie des Sciences.
  Paris}, 256:1887--1890, 1963.

\bibitem{Tele}
C.H. Bennett, G.~Brassard, C.~Crepeau, R.~Jozsa, A.~Peres, and W.K. Wootters.
\newblock {Teleporting an unknown quantum state via dual classical and
  Einstein-Podolsky-Rosen channels}.
\newblock {\em Physical Review Letters}, 70(13):1895--1899, 1993.

\bibitem{BvN}
G.~Birkhoff and J.~von Neumann.
\newblock The logic of quantum mechanics.
\newblock {\em Annals of Mathematics}, 37:823--843, 1936.

\bibitem{BoixoHeunen}
S.~Boixo and C.~Heunen.
\newblock Entangled and sequential quantum protocols with dephasing.
\newblock {\em Physical Review Letters}, 108:120402, 2012.

\bibitem{TheoryMine}
A.~Bundy, F.~Cavallo, L.~Dixon, M.~Johansson, and R.~McCasland.
\newblock The theory behind theorymine.

\bibitem{CarboniWalters}
A.~Carboni and R.~F.~C. Walters.
\newblock Cartesian bicategories {I}.
\newblock {\em Journal of Pure and Applied Algebra}, 49:11--32, 1987.

\bibitem{Chiri1}
G.~Chiribella, G.M. DÕAriano, and P.~Perinotti.
\newblock Probabilistic theories with purification.
\newblock {\em Physical Review A}, 81(6):062348, 2010.

\bibitem{Chiri2}
G.~Chiribella, G.M. DÕAriano, and P.~Perinotti.
\newblock Informational derivation of quantum theory.
\newblock {\em Physical Review A}, 84(1):012311, 2011.

\bibitem{Chomsky}
N.~Chomsky.
\newblock Tree models for the description of language.
\newblock {\em I.R.E. Transactions on Information Theory}, IT-2:113--124, 1956.

\bibitem{CCS}
S.~Clark, B.~Coecke, and M.~Sadrzadeh.
\newblock A compositional distributional model of meaning.
\newblock In {\em Proceedings of the Second Quantum Interaction Symposium
  (QI-2008)}, pages 133--140, 2008.

\bibitem{Kindergarten}
B.~Coecke.
\newblock Kindergarten quantum mechanics --- lecture notes.
\newblock In A.~Khrennikov, editor, {\em Quantum Theory: Reconsiderations of
  the Foundations III}, pages 81--98. AIP Press, 2005.
\newblock {a}rXiv:quant-ph/0510032.

\bibitem{AAAI}
B.~Coecke.
\newblock Automated quantum reasoning: Non logic - semi-logic - hyper-logic.
\newblock In {\em AAAI Spring Symposium: Quantum Interaction}, pages 31--38.
  AAAI, 2007.

\bibitem{SelingerAxiom}
B.~Coecke.
\newblock Axiomatic description of mixed states from {S}elinger's
  {CPM}-construction.
\newblock {\em Electronic Notes in Theoretical Computer Science}, 210:3--13,
  2008.

\bibitem{ContPhys}
B.~Coecke.
\newblock Quantum picturalism.
\newblock {\em Contemporary Physics}, 51:59--83, 2009.
\newblock {a}rXiv:0908.1787.

\bibitem{CD1}
B.~Coecke and R.~Duncan.
\newblock Interacting quantum observables.
\newblock In {\em Proceedings of the 37th International Colloquium on Automata,
  Languages and Programming (ICALP)}, Lecture Notes in Computer Science, 2008.

\bibitem{CD2}
B.~Coecke and R.~Duncan.
\newblock Interacting quantum observables: categorical algebra and
  diagrammatics.
\newblock {\em New Journal of Physics}, 13:043016, 2011.
\newblock {arXiv:quant-ph/09064725}.

\bibitem{CDKZ}
B.~Coecke, R.~Duncan, A.~Kissinger, and Q.~Wang.
\newblock {Strong complementarity and non-locality in categorical quantum
  mechanics}.
\newblock In G.~Chiribella and R.W. Spekkens, editors, {\em {Quantum Theory:
  Informational Foundations and Foils}}, page to appear. Springer, 2012.
\newblock {arXiv:1203.4988}.

\bibitem{CES}
B.~Coecke, B.~Edwards, and R.W. Spekkens.
\newblock Phase groups and the origin of non-locality for qubits.
\newblock {\em Electronic Notes in Theoretical Computer Science},
  270(2):15--36, 2011.
\newblock {a}rXiv:1003.5005.

\bibitem{CK}
B.~Coecke and A.~Kissinger.
\newblock {The compositional structure of multipartite quantum entanglement}.
\newblock In {\em Automata, Languages and Programming}, Lecture Notes in
  Computer Science, pages 297--308. Springer, 2010.
\newblock Extended version: {a}rXiv:1002.2540.

\bibitem{CMoore}
B.~Coecke and D.J. Moore.
\newblock Operational {G}alois adjunctions.
\newblock In B.~Coecke, D.J. Moore, and A.~Wilce, editors, {\em Current
  Research in Operational Quantum Logic: Algebras, Categories and Languages},
  volume 111 of {\em Fundamental Theories of Physics}, pages 195--218.
  Springer-Verlag, 2000.

\bibitem{CMW}
B.~Coecke, D.J. Moore, and A.~Wilce.
\newblock Operational quantum logic: An overview.
\newblock In B.~Coecke, D.J. Moore, and A.~Wilce, editors, {\em Current
  Research in Operational Quantum Logic: Algebras, Categories and Languages},
  volume 111 of {\em Fundamental Theories of Physics}, pages 1--36.
  Springer-Verlag, 2000.
\newblock arXiv:quant-ph/0008019.

\bibitem{CPaq}
B.~Coecke and {\'E}.~O. Paquette.
\newblock {POVM}s and {N}aimark's theorem without sums.
\newblock {\em Electronic Notes in Theoretical Computer Science}, 210:15--31,
  2008.
\newblock {arXiv:quant-ph/0608072}.

\bibitem{CatsII}
B.~Coecke and E.~O. Paquette.
\newblock Categories for the practicing physicist.
\newblock In B.~Coecke, editor, {\em New Structures for Physics}, Lecture Notes
  in Physics, pages 167--271. Springer, 2011.
\newblock {a}rXiv:0905.3010.

\bibitem{CPaqPav}
B.~Coecke, E.O. Paquette, and D.~Pavlovic.
\newblock {Classical and quantum structuralism}.
\newblock In S.~Gay and I.~Mackie, editors, {\em Semantic Techniques in Quantum
  Computation}, pages 29--69. Cambridge University Press, 2010.
\newblock {a}rXiv:0904.1997.

\bibitem{CPav}
B.~Coecke and D.~Pavlovic.
\newblock Quantum measurements without sums.
\newblock In G.~Chen, L.~Kauffman, and S.~Lamonaco, editors, {\em Mathematics
  of Quantum Computing and Technology}, pages 567--604. Taylor and Francis,
  2007.
\newblock {arXiv:quant-ph/0608035}.

\bibitem{CPV}
B.~Coecke, D.~Pavlovic, and J.~Vicary.
\newblock A new description of orthogonal bases.
\newblock {\em Mathematical Structures in Computer Science, to appear}, 2011.
\newblock {a}rXiv:quant-ph/0810.1037.

\bibitem{CPer}
B.~Coecke and S.~Perdrix.
\newblock Environment and classical channels in categorical quantum mechanics.
\newblock In {\em Proceedings of the 19th EACSL Annual Conference on Computer
  Science Logic (CSL)}, volume 6247 of {\em Lecture Notes in Computer Science},
  pages 230--244, 2010.
\newblock Extended version: {a}rXiv:1004.1598.

\bibitem{CSC}
B.~Coecke, M.~Sadrzadeh, and S.~Clark.
\newblock Mathematical foundations for a compositional distributional model of
  meaning.
\newblock {\em Linguistic Analysis}, 2010.

\bibitem{CSBayes}
B.~Coecke and R.W. Spekkens.
\newblock Picturing classical and quantum {B}ayesian inference.
\newblock {\em Synthese}, pages 1--46, 2011.
\newblock {arXiv:1102.2368}.

\bibitem{Dirac}
P.~A.~M. Dirac.
\newblock {\em The Principles of Quantum Mechanics (third edition)}.
\newblock Oxford University Press, 1947.

\bibitem{DD1}
L.~Dixon and R.~Duncan.
\newblock {Graphical reasoning in compact closed categories for quantum
  computation}.
\newblock {\em Annals of Mathematics and Artificial Intelligence},
  56(1):23--42, 2009.

\bibitem{quanto}
L.~Dixon, R.~Duncan, B.~Frot, A.~Merry, A.~Kissinger, and M.~Soloviev.
\newblock {\tt quantomatic}.
\newblock {h}ttp://dream.inf.ed.ac.uk/projects/quantomatic/, 2011.

\bibitem{DK}
L.~Dixon and A.~Kissinger.
\newblock Open graphs and monoidal theories.
\newblock {\em Mathematical Structures in Computer Science, to appear}, 2011.
\newblock {arXiv:1011.4114}.

\bibitem{RossThesis}
R.~Duncan.
\newblock Types for quantum computation, 2006.
\newblock DPhil Thesis, Oxford University.

\bibitem{DP2}
R.~Duncan and S.~Perdrix.
\newblock {Rewriting measurement-based quantum computations with generalised
  flow}.
\newblock In {\em Proceedings of {ICALP}}, Lecture Notes in Computer Science,
  pages 285--296. Springer, 2010.

\bibitem{EilenbergMacLane}
S.~Eilenberg and S.~Mac~Lane.
\newblock General theory of natural equivalences.
\newblock {\em Transactions of the American Mathematical Society}, 58(2):231,
  1945.

\bibitem{FMC}
Cl-.A. Faure, D.J. Moore, and C.~Piron.
\newblock Deterministic evolutions and {S}chr\"odinger flows.
\newblock {\em Helvetica Physica Acta}, 68(2):150--157, 1995.

\bibitem{Foulis}
D.~J. Foulis and C.~H. Randall.
\newblock Operational statistics. {I}. {Basic} concepts.
\newblock {\em Journal of Mathematical Physics}, 13(11):1667--1675, 1972.

\bibitem{Frege}
G.~Frege.
\newblock \"{U}ber sinn und bedeutung.
\newblock {\em Zeitschrift f\"ur Philosophie und Philosophische Kritik}, 1007,
  1892.

\bibitem{Girard}
J.Y. Girard.
\newblock {Linear logic}.
\newblock {\em Theoretical Computer Science}, 50(1):1--101, 1987.

\bibitem{GrefSadr}
E.~Grefenstette and M.~Sadrzadeh.
\newblock Experimental support for a categorical compositional distributional
  model of meaning.
\newblock In {\em EMNLP}, pages 1394--1404. ACL, 2011.

\bibitem{HalvorsonBook}
H.~Halvorson.
\newblock {\em Deep Beauty: Understanding the Quantum World Through
  Mathematical Innovation}.
\newblock Cambridge University Press, 2011.

\bibitem{HardyPicturalism}
L.~Hardy.
\newblock A formalism-local framework for general probabilistic theories
  including quantum theory.
\newblock 2010.
\newblock {arxiv:1005.5164}.

\bibitem{HasegawaHofmannPlotkin}
M.~Hasegawa, M.~Hofmann, and G.~D. Plotkin.
\newblock Finite dimensional vector spaces are complete for traced symmetric
  monoidal categories.
\newblock In A.~Avron, N.~Dershowitz, and A.~Rabinovich, editors, {\em Pillars
  of Computer Science}, volume 4800 of {\em Lecture Notes in Computer Science},
  pages 367--385. Springer, 2008.

\bibitem{Clare}
C.~Horsman.
\newblock Quantum picturalism for topological cluster-state computing.
\newblock {\em New Journal of Physics}, 13:095011, 2011.
\newblock {arXiv:1101.4722}.

\bibitem{IsaCosy}
M.~Johansson, L.~Dixon, and A.~Bundy.
\newblock Conjecture synthesis for inductive theories.
\newblock {\em Journal of Automated Reasoning}, 47(3):251--289, 2011.

\bibitem{JS}
A.~Joyal and R.~Street.
\newblock The geometry of tensor calculus {I}.
\newblock {\em Advances in Mathematics}, 88:55--112, 1991.

\bibitem{KellyLaplaza}
G.~M. Kelly and M.~L. Laplaza.
\newblock Coherence for compact closed categories.
\newblock {\em Journal of Pure and Applied Algebra}, 19:193--213, 1980.

\bibitem{QuantoCosy}
A.~Kissinger.
\newblock Synthesising graphical theories.
\newblock 2012.
\newblock {arXiv:1202.6079}.

\bibitem{KockBook}
J.~Kock.
\newblock {\em Frobenius algebras and 2D topological quantum field theories},
  volume~59.
\newblock Cambridge University Press, 2004.

\bibitem{Lack}
S.~Lack.
\newblock Composing {PROP}s.
\newblock {\em Theory and Applications of Categories}, 13:147--163, 2004.

\bibitem{Lambek0}
J.~Lambek.
\newblock The mathematics of sentence structure.
\newblock {\em American Mathematics Monthly}, 65, 1958.

\bibitem{Lambek1}
J.~Lambek.
\newblock Type grammar revisited.
\newblock {\em Logical Aspects of Computational Linguistics}, 1582, 1999.

\bibitem{LambekScott}
J.~Lambek and P.J. Scott.
\newblock {\em Introduction to higher order categorical logic}, volume~7.
\newblock Cambridge University Press, 1988.

\bibitem{MacLaneCoherence}
S.~Mac~Lane.
\newblock Natural associativity and commutativity.
\newblock {\em The Rice University Studies}, 49(4):28--46, 1963.

\bibitem{Mackey}
G.~M. Mackey.
\newblock {\em The mathematical foundations of quantum mechanics}.
\newblock W. A. Benjamin, New York, 1963.

\bibitem{DJMoore}
D.~J. Moore.
\newblock On state spaces and property lattices.
\newblock {\em Studies in History and Philosophy of Modern Physics},
  30(1):61--83, March 1999.

\bibitem{Penrose}
R.~Penrose.
\newblock Applications of negative dimensional tensors.
\newblock In {\em Combinatorial Mathematics and its Applications}, pages
  221--244. Academic Press, 1971.

\bibitem{Piron}
C.~Piron.
\newblock {\em Foundations of quantum physics}.
\newblock W. A. Benjamin, 1976.

\bibitem{PrelSadr}
A.~Preller and M.~Sadrzadeh.
\newblock Bell states and negative sentences in the distributed model of
  meaning.
\newblock {\em Electronic Notes in Theoretical Computer Science},
  270(2):141--153, 2011.

\bibitem{Redei}
M.~Redei.
\newblock Why {J}ohn von {N}eumann did not like the {H}ilbert space formalism
  of quantum mechanics (and what he liked instead).
\newblock {\em Studies in History and Philosophy of Modern Physics},
  27(4):493--510, 1996.

\bibitem{Schrodinger}
E.~Schr\"odinger.
\newblock Discussion of probability relations between separated systems.
\newblock {\em Cambridge Philosophical Society}, 31:555--563, 1935.

\bibitem{Schuetze}
H.~Sch{\"u}tze.
\newblock Automatic word sense discrimination.
\newblock {\em Computational linguistics}, 24(1):97--123, 1998.

\bibitem{Seely}
R.~A.~G. Seely.
\newblock Linear logic, {$*$}-autonomous categories and cofree algebras.
\newblock {\em Contemporary Mathematics}, 92:371--382, 1989.

\bibitem{SelingerCPM}
P.~Selinger.
\newblock Dagger compact closed categories and completely positive maps.
\newblock {\em Electronic Notes in Theoretical Computer Science}, 170:139--163,
  2007.

\bibitem{SelingerCompleteness}
P.~Selinger.
\newblock Finite dimensional {H}ilbert spaces are complete for dagger compact
  closed categories (extended abstract).
\newblock {\em Electronic Notes in Theoretical Computer Science},
  270(1):113--119, 2011.

\bibitem{Shor}
P.~W. Shor.
\newblock Polynomial-time algorithms for prime factorization and discrete
  logarithms on a quantum computer.
\newblock {\em SIAM Journal on Computing}, 26(5):1484--1509, 1997.

\bibitem{Wittgenstein}
L.~Wittgenstein.
\newblock {\em Philosophical Investigations}.
\newblock Basil \& Blackwell, 1972.

\bibitem{Swap}
M.~Zukowski, A.~Zeilinger, M.~A. Horne, and A.~K. Ekert.
\newblock {E}vent-ready-detectors {B}ell experiment via entanglement swapping.
\newblock {\em Physical Review Letters}, 71:4287--4290, 1993.

\end{thebibliography}

\end{document}